\documentclass[amsmath,amssymb,aps,prb,twocolumn]{revtex4}
\usepackage{latexsym}
\usepackage{graphicx}
\usepackage{times}
\usepackage{amsmath}
\usepackage{dcolumn}
\usepackage{subfigure}
\usepackage{latexsym,amsmath,amssymb,bm,euscript}
\def\MnScS{$\text{Mn}\text{Sc}_2\text{S}_4$\,} 
\def\CoAlO{$\text{Co}\text{Al}_2\text{O}_4$\,}
\def\ABX{$\text{A}\text{B}_2\text{X}_4$\,}

\newcommand{\be}{\begin{equation}}
\newcommand{\ee}{\end{equation}}

\begin{document}

\title{Theory of the Ordered Phase in A-site Antiferromagnetic Spinels}
\author{SungBin Lee}
\affiliation{Department of Physics, University of California, Santa Barbara, CA
93106-9530}

\author{Leon Balents} \affiliation{Kavli Insitute for Theoretical
  Physics, University of California, Santa Barbara, CA 93106-9530}

\date{\today} 

\begin{abstract} 
  Insulating spinel materials, with the chemical formula $AB_2X_4$,
  behave as diamond lattice antiferromagnets when only the A-site atom
  is magnetic.  Many exhibit classic signatures of frustration, induced
  not geometrically but by competing first and second neighbor exchange
  interactions.  In this paper, we further develop a
  theory\cite{Doron:order_by_disorder} of the magnetism of these
  materials, focusing on the physics observable within the ordered
  state.  We derive a phenomenological Landau theory that predicts the
  orientation of the spins within incommensurate spiral ordered states.
  It also describes how the spins reorient in a magnetic field, and how
  they may undergo a low temperature ``lock-in'' transition to a
  commensurate state.  We discuss microscopic mechanisms for these
  magnetic anisotropy effects.  The reduction of the ordered moment by
  quantum fluctuations is shown to be enhanced due to frustration.  Our
  results are compared to experiments on \MnScS, the best characterized
  of such A-site spinels, and more general implications are discussed.
  One prediction is that magnetically-induced ferroelectricity is
  generic in these materials, and a detailed description of the relation
  of the electric polarization to the magnetism is given.
\end{abstract}

\maketitle 


\section{Introduction}

Frustrated magnets, in which competing exchange interactions cannot be
simultaneously minimized, have long been a subject of theoretical and
experimental study.  Their fundamental interest comes from their
tendency to show more pronounced effects of fluctuations than their
unfrustrated counterparts, and from prospects of observing exotic ground
states as a consequence of frustration-induced sensitivity to weak
perturbations.  From a more practical materials science perspective,
they are of particular recent interest because they tend to display
non-collinear magnetic ordering. Such non-collinear ordering is quite
generally connected to magnetically-induced ferroelectricity, making
frustrated magnets a rich and productive hunting ground for
multiferroics.  In this paper, we study a particular class of frustrated
spinel materials, with the chemical formula \ABX, in which only the A
atom is magnetic.  Such materials are described as antiferromagnets on
the diamond lattice.  Somewhat surprisingly, although the diamond
lattice is not {\sl geometrically} frustrated and admits a simple two
sublattice collinear N\'eel state, many of these A-site magnetic spinels
do exhibit significant signs of frustration.  This includes a large
ratio (``frustration parameter'') $f= |\Theta_{CW}|/T_c$
\cite{ramirez94:_stron_geomet_frust_magnet} between the Curie-Weiss
temperature $\Theta_{CW}$ and an ordering or freezing temperature $T_c$.
For example, experiments find $f\approx
10-20$ in
\CoAlO\cite{tristan:geometric_frustration_in_the_cubic_spinels,suzuki:melting_of_antiferromagnetic_ordering}, and $f\approx 12$ in
\MnScS.\cite{fritsch:spin_orbital_frustration_in_MnScS}
A recent theoretical study attributes this to the competition between
first and second neighbor exchange interactions, $J_1$ and $J_2$, which
can be comparable in these materials.\cite{Doron:order_by_disorder}\
Theoretically, for $J_2/J_1> 1/8$, the classical ground state becomes
highly degenerate, consisting of coplanar spirals whose wavevector can
be {\sl arbitrarily} chosen on some ``spiral surface'' in momentum
space.  This degeneracy was suggested to be responsible for the observed
signs of frustration, including large $f$, prominent diffuse neutron
scattering in the paramagnetic state, and some low temperature specific
heat anomalies. While encouraging, many of the predictions of this
theory cannot currently be tested due to the absence of single crystal
neutron scattering data.

In this paper, we develop this theory further, in order both to capture
more detailed physical properties of this class of materials, and to
make further predictions which might more readily be compared to
existing and future experiments.  We focus on physics than can be
directly observed in the ordered state, which has been fairly well
characterized in \MnScS.  Specifically, we consider details of the {\sl
  magnetic anisotropy}, and the magnitude of the local ordered moments
at low temperature.  The theory of Ref.~\onlinecite{Doron:order_by_disorder}\
was based on an Heisenberg model, which possesses O(3) (or SU(2)) spin
rotation symmetry and hence exhibits no preference for the absolute
orientations of the spins themselves in the ordered state.
Experimentally, in \MnScS the spins are observed to lie in a definite
plane.  Moreover, the ordering wavevector describing the axis and pitch
of the spiral in real space displays a ``lock-in'' behavior at low
temperature, in which it becomes commensurate with the underlying spinel
lattice.  In the Heisenberg model, there is no explanation for this
lock-in.  We show here that both the choice of spiral plane and the
commensurate lock-in of the spiral wavevector can be understood by
considering magnetic anisotropy effects.  By an extended
phenomenological Landau analysis, we can describe the magnetic
orientation selection across the broader family of A-site spinels --
which has not yet been studied experimentally -- and predict some
interesting ``spin flop'' and reorientation effects in applied magnetic
fields.  We also consider, as mentioned, the value of the ordered
moment, which experimentally shows a relatively large (for an $S=5/2$
spin) $20\%$ suppression from the classical value.  We show that,
despite the large Mn$^{2+}$ spins, this can actually be accounted for by
quantum fluctuations, provided further neighbor interactions are
sufficiently small, due to the enhancement of fluctuations by
frustration.  Finally, we discuss the microscopic mechanisms behind the
magnetic anisotropy of these materials, which may arise both from
dipolar interactions and spin-orbit effects.  In \MnScS, we find that
spin-orbit induced exchange anisotropy is the only one of these two
mechanisms consistent with experimental observations.

We emphasize that though we pay particular attention to the comparison
with \MnScS, the A-site spinels comprise a quite large set of
interesting magnetic materials, and the theoretical analysis of this
paper is formulated in such a way as to apply to the entire family.  It
therefore has numerous implications for many materials, and should be
quite useful as a guide to future experiments.  Of particular interest
is the possibility of observing ferroelectricity and magnetoelectric
effects in these compounds.  Our modeling of magnetic anisotropy
contains the essential ingredients for a theory of magnetically-induced
ferroelectricity.  We present some basic observations of this type in
the Discussion at the end of the paper.

The remainder of the paper is organized as follows. In
Sec.~\ref{sec:spiral-spin-state}, we describe a phenomenological form of
the magnetic anisotropy in terms of the order parameter, based on
symmetry constraints, and the resulting ground states.  In
Sec.~\ref{sec:magn-proc}, we discuss the magnetization process and a
spin flop transition in a field.  Sec.~\ref{sec:comm-effects} discusses
tendency of the spiral wavevector to lock to commensurate values, and
associated phase transitions.  We show in
Sec.~\ref{sec:quantum-fluctuations} how quantum fluctuations can be
included in the theory.  Then, in Sec.~\ref{sec:micr-orig-magn} we
consider the possible microscopic sources of the magnetic anisotropy,
and conclude that in \MnScS, it is most likely dominated by spin-orbit
induced exchange anisotropy. We conclude in
Sec.~\ref{sec:discussion} with a summary of results, and a discussion of
experimental phenomena, including magnetically induced
ferroelectricity.  Some technical calculations are included in the
Appendices.

\section{Spiral spin state and spin rotational symmetry breaking}
\label{sec:spiral-spin-state}

%
\begin{figure}[t]
\vskip0.5cm
\scalebox{0.7}{\includegraphics{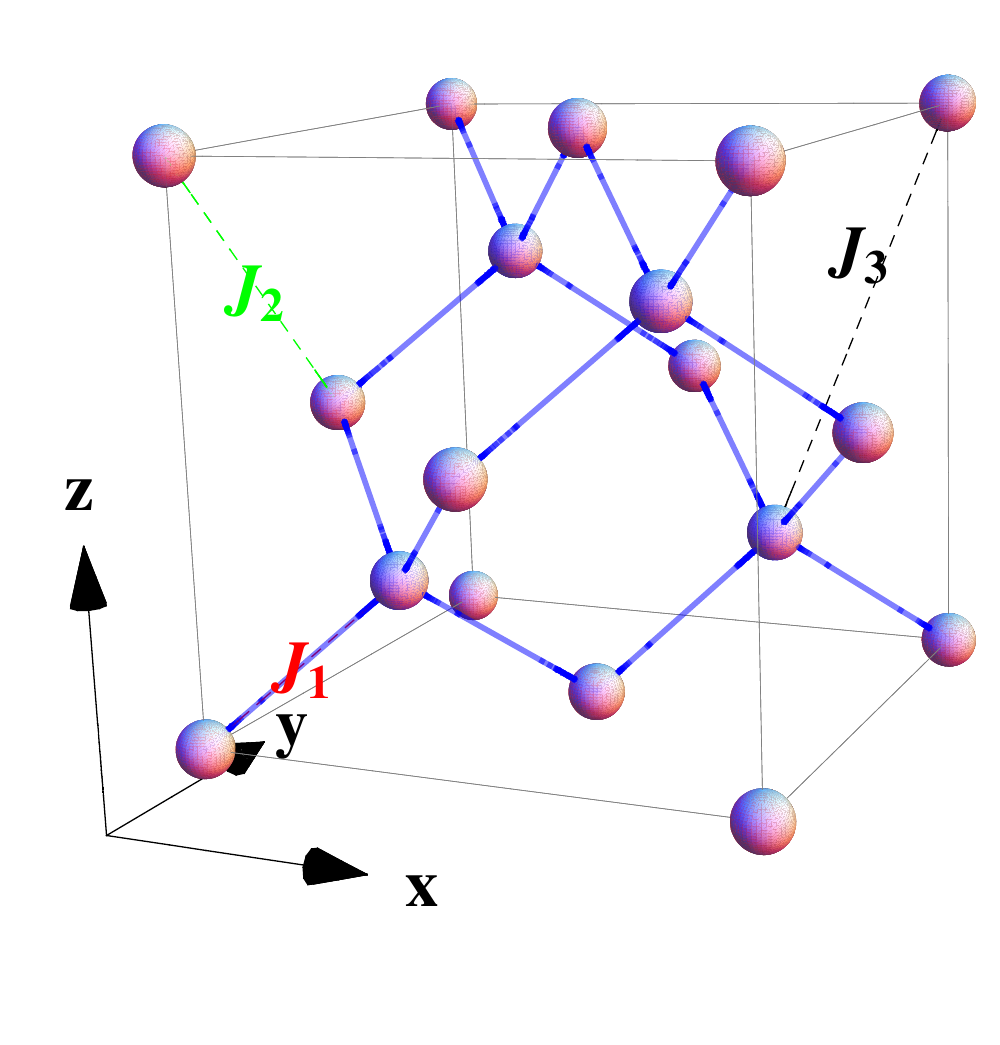}}
\caption{ The diamond lattice with the 1st, 2nd and 3rd nearest-neighbors coupling $J_1,J_2,J_3$ respectively.}
\label{diamond_lattice}
\end{figure}
%

\subsection{Heisenberg model and its ground states}
\label{sec:heisenberg-model-its}

A minimal Heisenberg model description for the magnetism of these
materials was studied in Ref.~\onlinecite{Doron:order_by_disorder}.  Here the
spins reside at the spinel A sites, which form a diamond lattice(see
Fig.\ref{diamond_lattice}), composed of the 2 interpenetrating fcc
lattice. The Hamiltonian, in zero magnetic field, is simply
\begin{equation}
  \label{eq:1}
  H_{\rm heis} = \frac{1}{2}\sum_{ij} J_{ij} {\bf S}_i \cdot {\bf S}_j.
\end{equation}
Here we consider classical unit vector spins $|{\bf S}_i|=1$.  We
consider coupling between up to third neighbor diamond sites,
i.e. $J_{ij}=J_1,J_2,J_3$ for first, second, and third neighbor sites,
respectively.  Though the diamond lattice with only nearest neighbor
spin exchange $J_1$ has an unfrustrated unique ground state, the
inclusion of additional interactions (2nd,3rd nearest neighbor etc)
rapidly produces frustration. Following the logic of
Ref.~\onlinecite{Doron:order_by_disorder}, we presume that the first second
nearest neighbor exchanges, $J_1,J_2$, are dominant, and treat the third
neighbor coupling $J_3$ as a small (but important) degeneracy breaking
perturbation. 

Ground states of this Hamiltonian can be found for arbitrary $J_i$ by
the method of Luttinger and Tisza.  They take the
form\cite{Doron:order_by_disorder} of coplanar spirals
\begin{equation}
   {\bf S}_i^{A(B)}=\dfrac{1}{2}{\bf d} e^{i{\bf k} {\cdot} {\bf x}_i \pm i\gamma/2}+c.c., \label{spiral_spin}\\
\end{equation}
where the order parameter ${\bf d}$ is a complex 3-component vector satisfying
\begin{eqnarray}
   {\bf d} \cdot {\bf d} & = &0, \label{eq:3}\\
   {\bf d} \cdot {\bf d}^* & = &2 .\nonumber
\end{eqnarray}
These two constraints, and the choice of $\gamma$, ensure that the
magnitude of each spin is unity, $|{\bf S}_i|=1$.  One has
\begin{equation}
  \label{eq:7}
  \gamma = {\rm Arg} \left[ {\sum_{i \in A, j \in B}}^{\!\!\!\prime} J_{ij}e^{i {\bf k}
      \cdot {\bf r}_{ij}}\right],
\end{equation}
where the sum $\sum_{}^\prime$ is taken over sites is taken over all sites
$j$ on the B sublattice, with $i$ fixed as an arbitrary A sublattice
site.  The physical meaning of ${\bf d}$ is made clear by solving the
constraints:
\begin{equation}
  \label{eq:4}
  {\bf d} = {\bf \hat e}_1 + i {\bf \hat e}_2,
\end{equation}
and defining
\begin{equation}
  \label{eq:5}
  {\bf \hat e}_3 = {\bf \hat e}_1 \times {\bf \hat e}_2 = \frac{i}{2}
  {\bf d}^{\vphantom*} \times {\bf d}^*.
\end{equation}
Here ${\bf\hat e}_1,{\bf\hat e}_2,{\bf\hat e}_3$ are three mutually
orthogonal unit vectors.  The first two span the plane in which the
spins reside, and ${\bf\hat e}_3$ is the unique normal to the plane.
A phase rotation of ${\bf d}$ rotates the spins within the plane, or
equivalently translates the spiral along its axis, while leaving the
spin plane and hence ${\bf\hat e}_3$ unchanged.

\begin{figure}[t]
\vskip0.5cm
  \includegraphics[width=0.45\textwidth]{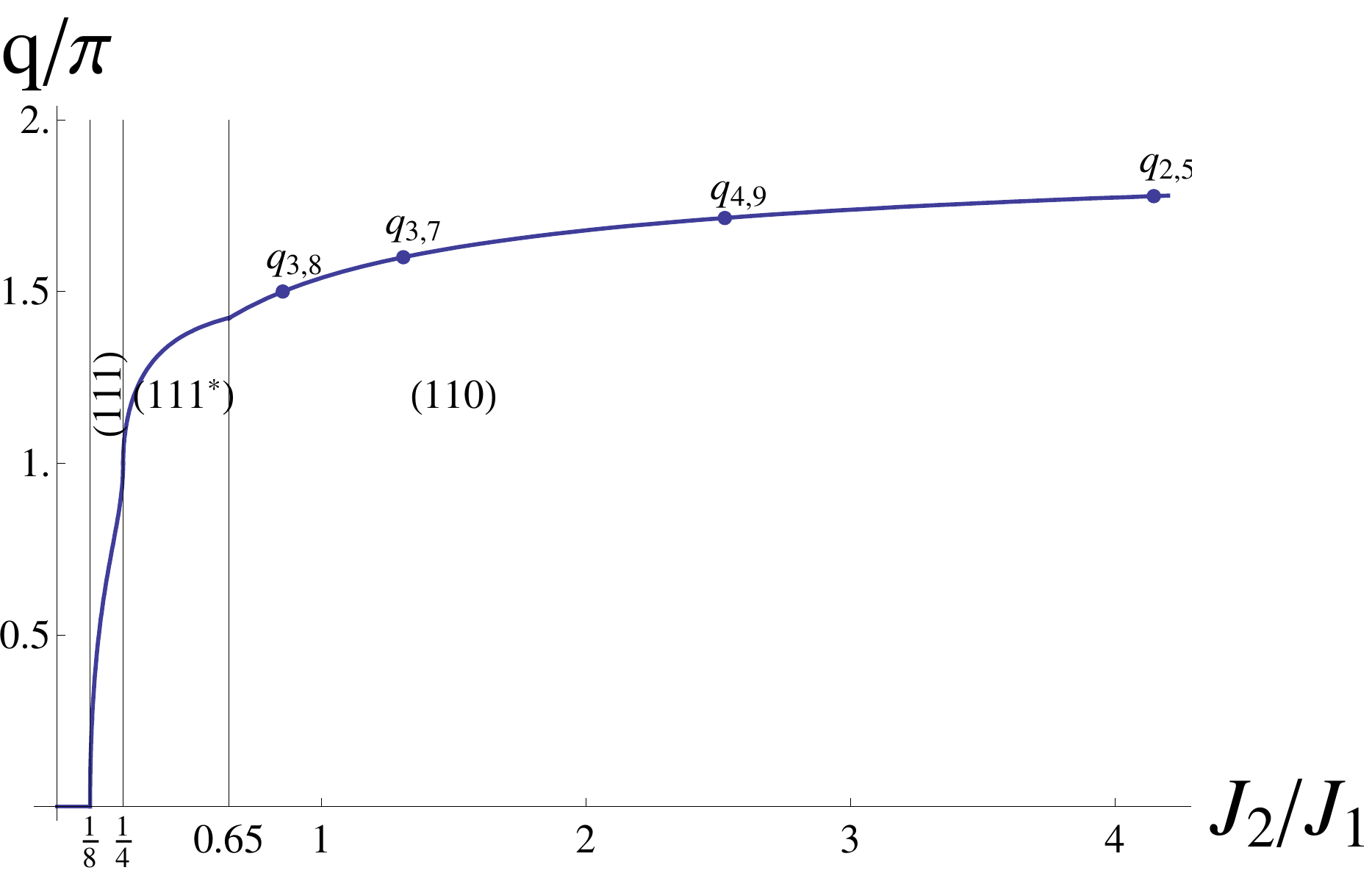}
  \caption{The selected wavevector of the diamond antiferromagnet for
    antiferromagnetic $J_3$.  We plot $q/\pi$ as a function of
    $J_2/J_1$, where the ground state wavevector has the form $(q,q,k)$.
    The direction (choice of $k$) is indicated by the labels
    (111),(111$^*$), (110) -- see text and
    Appendix~\ref{sec:splitt-spir-surf} for details -- in each of the
    regions separated by vertical lines.  The first four lowest order
    commensurate wavevectors $q_{m,n}$ for which lock-in transitions are
    expected are also indicated by labeled dots (see
    Sec.~\ref{sec:comm-effects}). }
\label{preferred_wave_vector} 
\end{figure}

The energy of spiral states of this type is readily evaluated.  It is
sufficient to linearize in $J_3$, in which case one finds the energy
per unit cell (this is twice the energy per spin)
\begin{equation}
  \label{eq:14}
  E_J({\bf k}) = E_{12}({\bf k})+E_3({\bf k}),
\end{equation}
where $E_{12}$ and $E_3$ are the contributions from the
large $J_1,J_2$ exchanges and the smaller $J_3$ exchange, respectively.
Explicitly, 
\begin{eqnarray}
   E_{12}&=& 16J_2\left(\Lambda({\bf k})-\frac{|J_1|}{8J_2}\right)^2-4J_2-\dfrac{J_1^2}{4J_2}, \\ 
\delta E_{3} &=& J_3 \frac{\Sigma({\bf k})}{\Lambda({\bf k})},
\end{eqnarray} 
\begin{widetext}
with
\begin{eqnarray}
\label{eq:6}
  && \Lambda({\bf k}) = 
    \Big[ \cos^2\frac{k_x}{4} \cos^2\frac{k_y}{4} \cos^2\frac{k_z}{4} 
  + \sin^2\frac{k_x}{4} \sin^2\frac{k_y}{4}
  \sin^2\frac{k_z}{4}\Big]^{1/2}, \nonumber  \\
  &&   \Sigma({\bf k}) = \cos k_x \left(1+2 \cos
  \frac{k_y}{2} \cos \frac{k_z}{2}\right) + 2 \cos
  \frac{k_x}{2} \cos \frac{k_y}{2} + \mbox{cyclic perms}.
\end{eqnarray}
\end{widetext}
Treating $J_3$ perturbatively, we first minimize $E_{12}$.  For
$J_2/|J_1|<1/8$, the minimum occurs for ${\bf k=0}$, while for
$J_2/|J_1|>1/8$, it occurs along the surface defined by $\Lambda({\bf
  k})= |J_1|/{8J_2}$. In the latter case, the 3rd nearest-neighbor
exchange breaks the ``spiral surface'' degeneracy.  A combination of
analytical and numerical arguments (see
Appendix~\ref{sec:splitt-spir-surf}) determine the selected wavevectors
on the spiral surface.  We assume antiferromagnetic $J_3>0$, in which
case the minimum energy is realized with a wave vector of the form ${\bf
  q}=(q,q,k)$, where the relation of $k$ to $q$ varies depending upon
the magnitude of $J_2/J_1$.  The direction of the wavevector thereby
varies from the (111) to the (110) directions, with an intermediate
(111$^*$) region in which the $k$ is chosen as close as possible to $q$,
since the (111) directions do not intersect the spiral surface.  See
Fig.\ref{preferred_wave_vector}\ and Appendix~\ref{sec:splitt-spir-surf}
for further details.  We note that this wavevector, determined from the
third nearest-neighbor exchange $J_3$, is different with the one
determined by thermal fluctuations.\cite{Doron:order_by_disorder}

For the specific material \MnScS, the magnetic structure is known from
neutron diffraction.\cite{krimmel:magnetic_ordering}  At low temperature
the ordering wavevector is ${\bf k}={\bf q} \equiv 3\pi/2(1,1,0)$, and
the refinement indicates ferromagnetic $J_1<0$.  Comparison to the
theoretical structure and the measured Curie-Weiss temperature allows
one to constraint the couplings\cite{Doron:order_by_disorder}.  When
$J_3$ is very small, one has $J_1
\simeq -10.5K$ and $J_2 \simeq 8.75K$.   More generally, fixing ${\bf
  k}=3\pi/2(1,1,0)$, one has
\begin{eqnarray}
\label{eq:J2J3}
 J_3/|J_1|=\frac{-1+(4-2\sqrt{2})J_2/|J_1|}{4\sqrt{2}-3}.
\end{eqnarray} 
From this relation, $J_2/|J_1|$ varies from $0.88$ to $0.94$ when
$J_3/|J_1|$ is increased from $0.01$ to $0.04$.

%
%

\subsection{Magnetic anisotropy}
\label{sec:magnetic-anisotropy}

The Heisenberg model leaves the plane and phase of the spin spiral
undetermined, because they can be continuously rotated using the SU(2)
symmetry of the Hamiltonian.  In reality, this symmetry is broken by the
crystal lattice and ``spin orbit'' effects (in fact arising both from
quantum mechanical spin-orbit coupling and dipolar interactions between
spins) that couple spin and spatial rotations.  Indeed, in \MnScS, it is
known that the spins in the $(110)$ spiral lie in a $(001)$ plane.  This
is determined by physics outside the Heisenberg model.  Furthermore, the
commensurate {\sl magnitude} of the wavevector -- ${\bf q}=q_0 (1,1,0)$
with $q_0=3\pi/2$ {\sl exactly} within experimental resolution -- is
also related to anisotropy effects.  In the Heisenberg model, obtaining
this value of $q_0$ at $T=0$ requires fine-tuning of the ratio of
$J_2/J_1$, and even with such tuning, the magnitude would generally
deviate at $T>0$.

To understand these effects, we first adopt a phenomenological Landau
theoretic approach constrained only by symmetry.  This consists of
time-reversal invariance, which reverses spins, and the space group,
Fd$\bar{\rm 3}$m, of the spinel lattice.  The full space group is
generated by 6 operations, which may be expressed in terms of
translations ${\mathcal T}_{\bf t}$ by the vector ${\bf t}$, rotations
${\mathcal R}_{\bf n}[\theta]$ by angle $\theta$ about the ${\bf n}$
axis, and the inversion ${\mathcal I}$ about the origin.  In our
coordinate system, the generators ${\mathcal G}_i$ are 
\begin{eqnarray}
 {\mathcal G}_1 & = & {\mathcal T}_{\frac{3}{4},\frac{1}{4},\frac{1}{2}}
 \circ {\mathcal R}_{001}[\pi], \\
 {\mathcal G}_2 & = & {\mathcal T}_{\frac{1}{4},\frac{1}{2},\frac{3}{4}}
 \circ {\mathcal R}_{010}[\pi], \\
 {\mathcal G}_3 & = & {\mathcal R}_{111}[\frac{2\pi}{3}], \\
 {\mathcal G}_4 & = & {\mathcal T}_{\frac{3}{4},\frac{1}{4},\frac{1}{2}}
 \circ {\mathcal R}_{110}[\pi], \\
 {\mathcal G}_5 & = & {\mathcal I}, \\
{\mathcal G}_6 & = & {\mathcal T}_{0,\frac{1}{2},\frac{1}{2}}.
 \label{symmetries}
\end{eqnarray}
Because the spin is a pseudo-vector, its transformation under each of
these operations is given by
\begin{equation}
{\bf S}({\bf x}) \rightarrow Det[\hat{{\bf O}}] \cdot \hat{{\bf O}}^{-1}
\cdot {\bf S}(\hat{{\bf O}} \cdot {\bf r}+{\bf t}),
 \label{spin_transform}
\end{equation}
where $\hat{\bf O}$ is the orthogonal matrix giving the
rotation/inversion part of the operation (${\bf r}\rightarrow \hat{\bf
  O}\cdot {\bf r}$) and ${\bf t}$ is the translation vector.  

We are interested in the effect of spin-orbit coupling {\sl within} the
ordered phase of these materials.  In this case, the symmetry is already
reduced from that of the full crystal by the magnetic order.
Specifically, we assume an ordered state of the form predicted by the
Heisenberg model, i.e. satisfying Eq.~(\ref{spiral_spin}) with ${\bf k}$
determined to be one of the values selected by $J_1,J_2,J_3$, but with
${\bf d}$ arbitrary up to the constraints in Eq.~(\ref{eq:3}).  We seek
a Landau free energy as a function of ${\bf d}$.  Since we restrict to a
fixed ${\bf k}$, we should consider only those symmetry operations which
leave ${\bf k}$ invariant (up to inversion).  This is the {\sl little
  group} of the wavevector ${\bf k}$.  Under each element in this little
group, because the wavevector is invariant, one can define a
corresponding transformation for ${\bf d}$, under which the free energy
must be invariant.

We consider the two major regimes of phase space in which the form of
${\bf k}$ is simple.  For $1/8<J_2/|J_1|<1/4$, we have ${\bf k}=k(1,1,1)$.
The little group is generated by the transformations ${\mathcal G}_3,
{\mathcal G}_5, {\mathcal G}_6$ in this case.  Under these operations,
the order parameter transforms according to
\begin{eqnarray}
  \label{eq:2}
  {\mathcal G}_3: d_1 & \rightarrow & d_3,\qquad d_2\rightarrow d_1,
  \qquad d_3 \rightarrow d_2, \nonumber \\
  {\mathcal G}_5: {\bf d} & \rightarrow & {\bf d}^*, \nonumber \\
  {\mathcal G}_6: {\bf d} & \rightarrow & e^{ik} {\bf d}.   
\end{eqnarray}
In the case $J_2/J_1\gtrsim 0.7$, one has ${\bf k}=k(1,1,0)$, for which the
little group is generated instead by ${\mathcal G}_1,{\mathcal G}_4,
{\mathcal G}_5, {\mathcal G}_6$.  Under these operations, we find
\begin{eqnarray}
  \label{eq:8}
   {\mathcal G}_1: {\bf d} & \rightarrow & e^{-ik} {\bf d}^*, \nonumber \\
   {\mathcal G}_4: {\bf d} & \rightarrow & e^{ik}
   \left(\begin{array}{ccc} 0 & 1 & 0 \\ 1 & 0 & 0 \\ 0 & 0 &
       1 \end{array}\right) {\bf d},  \nonumber\\
  {\mathcal G}_5: {\bf d} & \rightarrow & {\bf d}^*, \nonumber \\
 {\mathcal G}_6: {\bf d} & \rightarrow & e^{ik/2} {\bf d}.  
\end{eqnarray}

Using these symmetries, we can determine the most general allowed form
of the free energy at any given order in ${\bf d}$, for each of these
two cases.  Our focus is on terms which violate $SU(2)$ symmetry,
induced by spin-orbit coupling or dipolar interactions.  As usual within
Landau theory, we expect terms which involve smaller powers of the order
parameter to be most important.  We therefore consider the leading {\sl
  quadratic} terms  other than the trivial $|{\bf d}|^2$ one.  For
the ${\bf k}=(k,k,k)$ states, we find a single non-trivial invariant:
\begin{equation}
 f_{111}({\bf d}) \equiv c\left[d_3^*(d^{\vphantom*}_1+d^{\vphantom*}_2)+d_2^*(d^{\vphantom*}_1+d^{\vphantom*}_3)+d_1^*(d^{\vphantom*}_2+d^{\vphantom*}_3)\right].
 \label{non_vanishng_111}
\end{equation}
For the wave vector k(1,1,0), the quadratic free energy contains two
non-trivial invariants:
\begin{equation}
 f_{110}({\bf d}) \equiv c_1(d^*_1d_2^{\vphantom*}+{\rm
   c.c})+c_2d^*_3d_3^{\vphantom*}. 
 \label{non_vanishng_110}
\end{equation}

%
\begin{figure}[t]
\vskip0.5cm
  \includegraphics[width=0.45\textwidth]{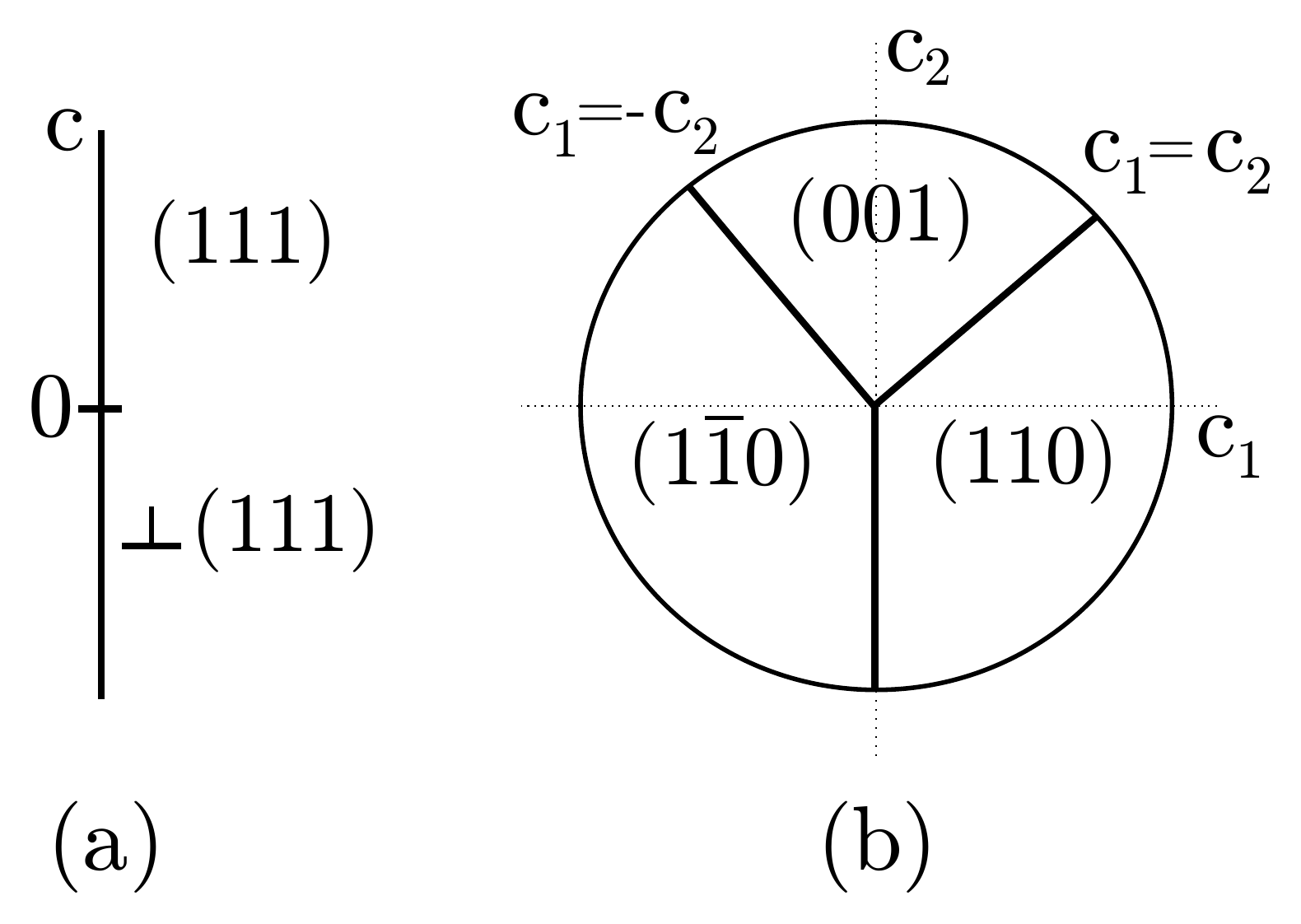}
  \caption{Directions of the normal ${\bf \hat e}_3$ to the plane of
    spin ordering selected by magnetic anisotropy terms in the cases (a)
    of a (111) wavevector and (b) of a (110) wavevector.  In (a), the
    symbol $\perp$(111) indicates that any plane with ${\bf\hat
      e}_3\cdot (111)=0$ is a ground state.}
\label{c1-c2_phase_diagram}
\end{figure}
%

These quadratic terms distinguish different planes in which the spins
spiral energetically.  We note that  both $f_{111}$ and $f_{110}$ are
invariant under arbitrary phase rotations of the ${\bf d}$ fields.
Physically, this implies rotations of the vectors ${\bf \hat e}_1$ and ${\bf
 \hat e}_2$ within the plane normal to ${\bf\hat e}_3$ cost no energy.
Therefore we expect that these terms may be rewritten in terms of ${\bf
  \hat e}_3$ alone.  This is indeed the case.  To do so, it is
convenient to introduce a parametrization of ${\bf d}$ which solves the
constraints in Eqs.~(\ref{eq:3}):
\begin{equation}
  \label{eq:9}
  {\bf d} = z_\alpha \epsilon_{\alpha\beta} {\boldsymbol
    \sigma}_{\beta\gamma} z_\gamma,
\end{equation}
where we have defined the spinor $z_\alpha$,
\begin{equation}
  \label{eq:10}
  z  = (e^{i\phi_1}\cos{\theta} , e^{i\phi_2}\sin{\theta}),
\end{equation}
which satisfies $|z_1|^2+|z_2|^2 = 1$.  Here ${\boldsymbol\sigma}$ is
the vector of Pauli matrices, and $\epsilon_{\alpha\beta}$ is the
anti-symmetric matrix with $\epsilon_{12}=1$.  It is straightforward to
show that
\begin{equation}
  \label{eq:74}
  {\bf\hat e}_3 = z_\alpha^* {\boldsymbol\sigma}_{\alpha\beta} z_\beta^{\vphantom*}.
\end{equation}
By explicit evaluation using Eqs.~(\ref{eq:9},\ref{eq:10}), one can
readily show
\begin{eqnarray}
  \label{eq:11}
  f_{111} & = & c\left(1 - \left[ e_3^x + e_3^y + e_3^z\right]^2\right), \\
  \label{eq:24} f_{110} & = & -2 c_1 e_3^x e_3^y + c_2 \left[ (e_3^x)^2 + (e_3^y)^2\right]
\end{eqnarray}

Now the energetically preferred plane for the spins is apparent.  They
are illustrated in Fig.~\ref{c1-c2_phase_diagram}. For ${\bf
  k}=(k,k,k)$, the ground state has ${\bf \hat e}_3= (1,1,1)/\sqrt{3}$
for $c>0$, and ${\bf \hat e}_3\cdot (1,1,1)=0$ for $c<0$ (i.e. in the
latter case, the vector ${\bf \hat e}_3$ is still free to rotate
anywhere within a plane).  For ${\bf k}=(k,k,0)$, three distinct
directions of ${\bf \hat e}_3$ are possible depending upon the values of
$c_1,c_2$ -- see Fig.~\ref{c1-c2_phase_diagram} for details.

At this stage it is possible to compare with experimental results on
\MnScS.  Refined neutron scattering data in
Ref.~[\onlinecite{krimmel:magnetic_ordering}] indicated spiral order of
the type discussed here with wavevector ${\bf q}=(q,q,0)$ and spins
aligned within the $(001)$ plane. We see that the Landau theory indeed
captures this order, provided the phenomenological parameters $c_1,c_2$
are taken to lie within region I of the phase diagram in
Fig.~\ref{c1-c2_phase_diagram}.  Note that this is not ``fine-tuning'',
as this region occupies a finite fraction of the phase diagram.
However, it is still interesting to understand the microscopic reason
for the system to be in region I rather than II or III.  We will return
to this question in Sec\ref{sec:micr-orig-magn}.

\section{Magnetization process}
\label{sec:magn-proc}

In this section, we consider the evolution of the spin state in an
applied magnetic field.  Neglecting magnetic anisotropy, we may expect a
smooth evolution, in which the spins adopt a canted (conical)
configuration with a non-vanishing component along the field, and this
canting gradually increases until the spins become fully aligned at
saturation.  In the presence of magnetic anisotropy, however, the spins
have an intrinsic preference for particular planes, which, in some field
orientations, competes with the tendency of the spins to adapt to the
field. We study these two situations below.

\subsection{Heisenberg model}
\label{sec:heisenberg-model}

We first neglect magnetic anisotropy and consider simply the classical
Heisenberg Hamiltonian with an added Zeeman magnetic field
\begin{equation}
 H_{J,h}= \dfrac{1}{2}\sum_{i, j} J_{ij}{\bf S}_i \cdot {\bf S}_j
 -\sum_{i}{\bf h} \cdot {\bf S}_i .
  \label{H_mag}
\end{equation}
We seek ground states with normalized spins $|{\bf S}_i|=1$, using
following ansatz:
\begin{equation}
  \label{eq:12}
  {\bf S}_i^{A(B)} = \frac{1}{2}{\bf d} e^{i({\bf k} {\cdot} {\bf x}_i \pm
    \gamma/2)}+c.c + {\bf m},  
\end{equation}
with the constraints
\begin{eqnarray}
  \label{eq:13}
  {\bf d} \cdot {\bf d} & = & 0, \\
 \label{eq:15} {\bf d}  \cdot  {\bf m} & = & 0,  \\
 \label{eq:16} \frac{1}{2}{\bf d} \cdot {\bf d}^*+
  {\bf m}^2 & = & 1.                                  
\end{eqnarray}
We now evaluate the energy for these states.  It is necessary to
consider ferromagnetic and antiferromagnetic $J_1$ separately.  
\begin{figure}[t]
  \centering
  \includegraphics[width=0.45\textwidth]{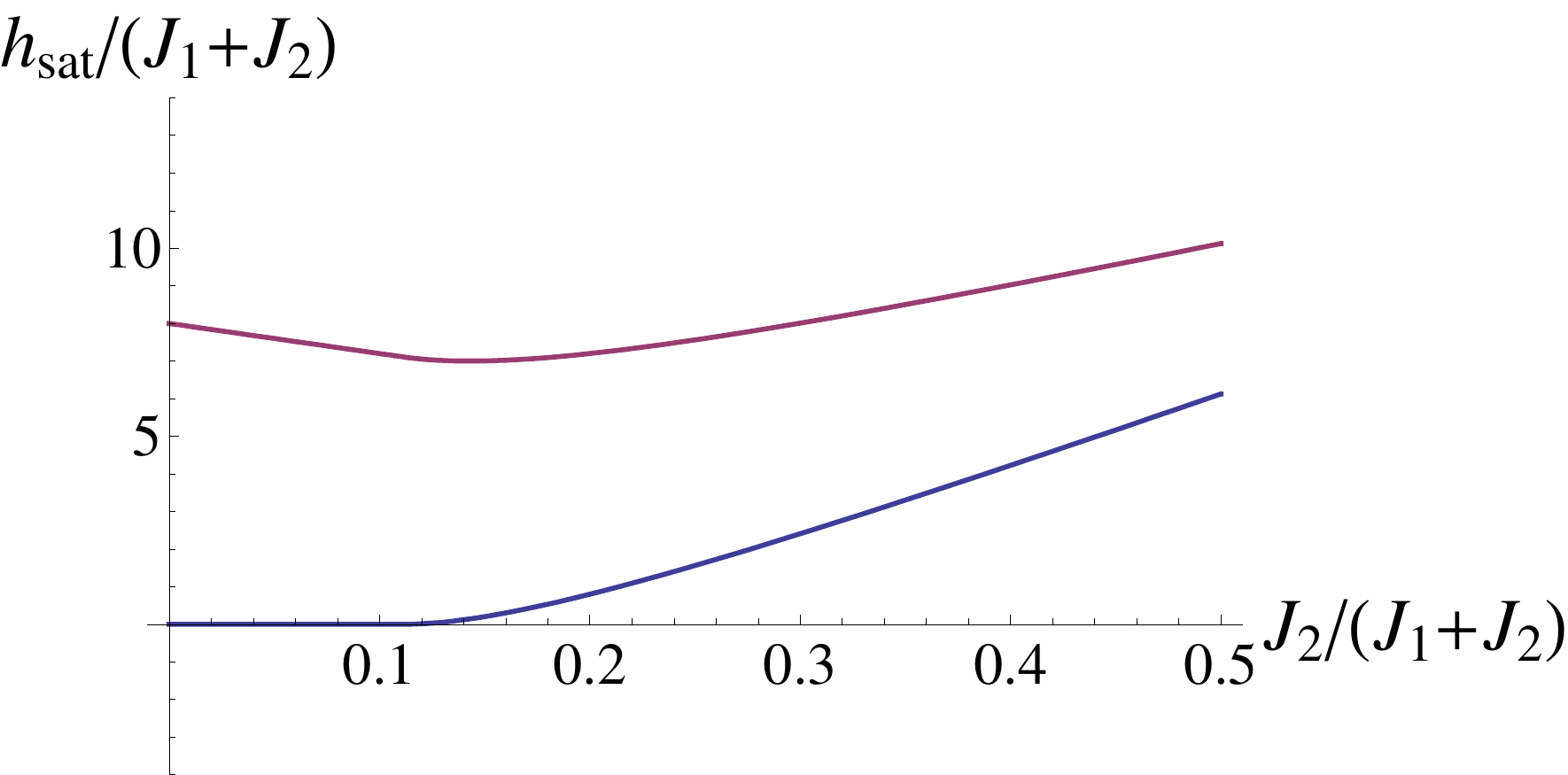}
  \caption{Saturation fields $h_{sat}/(J_1+J_2)$ as a function of
    $J_2/(J_1+J_2)$, for the ferromagnetic (lower curve) and
    antiferromagnetic (upper curve) cases.  } 
  \label{fig:sat}
\end{figure}
\subsubsection{Ferromagnetic $J_1$}
\label{sec:ferromagnetic-j_1}

In the ferromagnetic case, evaluating the energy per unit cell using the
Hamiltonian in Eq.~(\ref{H_mag}), one obtains
\begin{equation}
  E^{FM}_{J,h}=\frac{1}{2} E_J({\bf k}) |{\bf d}|^2  + |{\bf m}|^2
  E_J({\bf 0}) - 2{\bf h} \cdot {\bf m}.
 \label{H_spin_magnetization}
\end{equation}
Here $E_J({\bf k})$ is the energy function for a pure spiral in zero
field, given in Eq.~(\ref{eq:14}).  

This energy function is minimized as follows.  Only the third term is
dependent upon the orientation of ${\bf d}$ and ${\bf m}$, and it is
minimized if we choose ${\bf m}=m {\bf \hat h}$ along the field
direction.  Then we must choose, similarly to Eq.~(\ref{eq:4}),
\begin{equation}
  \label{eq:17}
  {\bf d}= \sqrt{1-m^2} ({\bf\hat e}_1+ i {\bf \hat e}_2),
\end{equation}
with ${\bf \hat e}_3= {\bf \hat e}_1\times {\bf \hat e}_2 = {\bf \hat
  h}$.  This indeed describes a conical spin state.  For fixed $m$ and
hence $|{\bf d}|^2=2(1-m^2)$, the energy is minimized by the wavevector
${\bf k}={\bf q}$ which minimized $E_J({\bf k})$.  This implies that the
wavevector is independent of magnetic field.  Finally, we can minimize
over $m$, which gives
\begin{equation}
  \label{eq:18}
  m = \frac{h}{h_{sat}},
\end{equation}
which is valid for fields below the saturation field, which in this
ferromagnetic case is
\begin{equation}
  \label{eq:19}
  h^{FM}_{sat}= E_J({\bf 0}) - E_J({\bf q}) \equiv \Delta E.
\end{equation}
Here we define $\Delta E$ for later convenience.
We see that the magnetization increases perfectly linearly up to saturation.
The saturation field itself varies with the exchange couplings and in
particular $J_2/J_1$ in a non-trivial manner as the ordering wavevector
${\bf q}$ varies -- see Fig.~\ref{fig:sat}.  Since the ground state
itself is ferromagnetic for $J_2<J_1/8$, the saturation field vanishes
in this region.

\subsubsection{Antiferromagnetic $J_1$}
\label{sec:antif-j_1}

Next consider the case of antiferromagnetic $J_1$.  In this case, the
energy function is
\begin{equation}
  \label{eq:20}
   E^{AFM}_{J,h}=\frac{1}{2} E_J({\bf k}) |{\bf d}|^2  + (E_J({\bf
     0})+8 J_1) |{\bf m}|^2
   -2 {\bf h} \cdot {\bf m}.
\end{equation}
The difference from Eq.~(\ref{H_spin_magnetization}) can be understood
as arising because of the cost $8J_1$ of flipping the four
nearest-neighbor bonds per site from anti-parallel to parallel spin
alignment.  Repeating the analysis of the previous subsubsection, we
again find a linear magnetization curve (i.e. Eq.~(\ref{eq:18})), but
with the saturation field
\begin{equation}
  \label{eq:21}
    h^{AFM}_{sat}= 8J_1 + \Delta E.
\end{equation}

\subsection{Anisotropy and spin flop transition}
\label{sec:anisotropic-case}

We now turn to the effects of magnetic anisotropy, and in particular the
competition between the magnetic field and the intrinsic preference for
the spin ordering plane.  Lacking a microscopic model for the
anisotropy, we cannot reliably explore the full phase diagram for all
fields.  However, since we expect that the anisotropy is relatively weak
compared to the exchange, the portion of phase space in which the field
and anisotropy are actually competitive is restricted to small fields.
In this regime, the contribution of the anisotropy to the energy should
be approximately unchanged from that at zero field, and hence we may
model it by the {\sl same} phenomenological function given in
Sec.~\ref{sec:magnetic-anisotropy}.  That is, we add to the Heisenberg
energy $E_{J,h}$ the terms $f_{111},f_{110}$, as appropriate.  

Let us focus on ferromagnetic $J_1$ with ${\bf q}=(q,q,0)$ for
simplicity.  The discussion is not significantly modified in the
antiferromagnetic case.  The energy function is now
\begin{equation}
  \label{eq:22}
  E_{tot}^{FM} =  \frac{1}{2} E_J({\bf k}) |{\bf d}|^2  + |{\bf m}|^2
  E_J({\bf 0}) - 2{\bf h} \cdot {\bf m} + f_{110}[{\bf d}].
\end{equation}
In small fields, we may fix ${\bf k}={\bf q}$ the zero-field ordering
wavevector which minimized $E_J$.  We can use Eq.~(\ref{eq:17}), with
however ${\bf \hat e}_3 = {\bf \hat m}$ not necessarily parallel to
${\bf h}$.  Inserting this into the energy, we find
\begin{eqnarray}
  \label{eq:23}
  E_{tot}^{FM} &  = &  E_J({\bf q})  + \Delta E |{\bf m}|^2
  - 2{\bf h} \cdot {\bf m} \nonumber \\
  & & + f_{110}[{\bf \hat e}_3={\bf \hat m}].
\end{eqnarray}
where $f_{110}[{\bf\hat e}_3]$ is given in Eq.~(\ref{eq:24}).  Here we
have approximated $m\approx 0$ in the anisotropy term, since the
neglected corrections are of $O(m^2 c_{1,2})$, i.e. small both in the
magnetization and the anisotropy.  

We can now minimize Eq.~(\ref{eq:23}) over the magnitude of the
magnetization at fixed orientation, which gives
\begin{equation}
  \label{eq:25}
  m = \frac{{\bf h}\cdot {\bf\hat m}}{      \Delta E},
\end{equation}
and the energy, which now depends only upon the orientation ${\bf \hat
  m}$:
\begin{equation}
  \label{eq:26}
  E_{tot}^{FM}({\bf \hat m}) = -\frac{({\bf h}\cdot {\bf \hat
      m})^2}{\Delta E} + f_{110}[{\bf \hat m}]
\end{equation}
up to constants independent of ${\bf \hat m}$.  We caution that in these
expressions, it is possible to take ${\bf\hat m}\cdot{\bf h}=0$, in
which case the actual magnetization vanishes, but ${\bf \hat m}={\bf\hat
  e}_3$ still defines the plane of the spiral.

\begin{figure}[hbtp]
  \centering
  \includegraphics[width=0.45\textwidth]{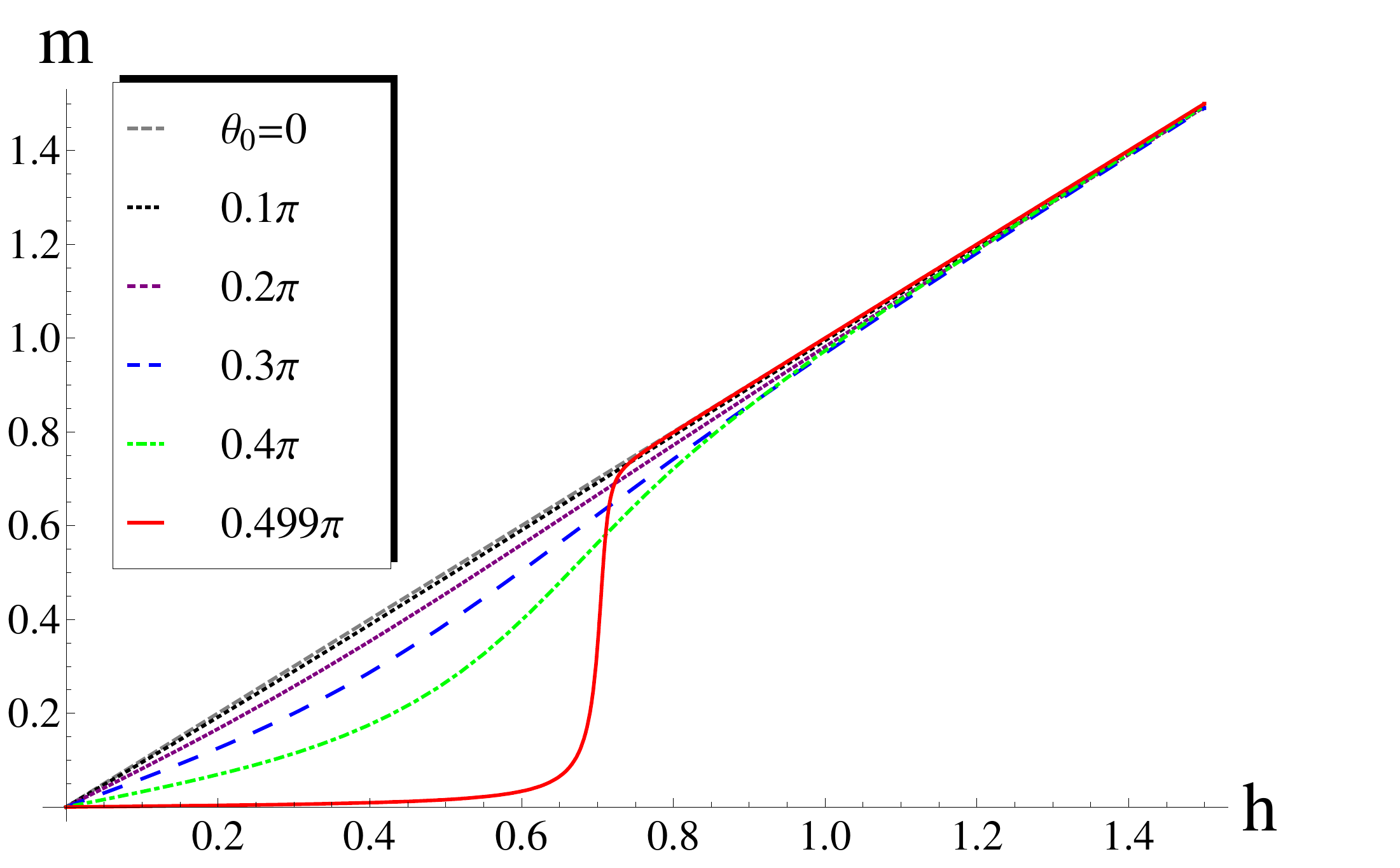}
  \caption{Magnitude of the magnetization $m$ versus dimensionless field
    ${\sf h}$ for $\theta_h=0,0.1\pi,0.2\pi,0.3\pi,0.4\pi,
    0.499\pi$ (from the top curve to the  bottom curve).}
  \label{fig:magcurves}
\end{figure}

To determine ${\bf \hat m}$, we must minimize Eq.~(\ref{eq:26}). Let us
first consider the special case $c_1=0, c_2>0$.  Then we may presume
${\bf\hat m}$ lies in the plane spanned by ${\bf\hat z}$ and ${\bf \hat
  h}$.  Taking the angle of ${\bf \hat m}$ with the $z$ axis as
$\theta$, and the angle of ${\bf \hat h}$ with the $z$ axis as
$\theta_h$, the energy is
\begin{eqnarray}
  \label{eq:27}
  E_{tot}^{FM} & = &  -c_2\cos^2\theta - \frac{h^2}{\Delta E}
  \cos^2(\theta-\theta_h), \\
  & = & - A \cos[2(\theta-\theta_0)] + {\rm const.},
\end{eqnarray}
where
\begin{eqnarray}
  \label{eq:29}
  A & = & \frac{c_2}{4}\sqrt{1+ 4{\sf h}^2 + 4 {\sf h} \cos 2\theta_h}, \\
  \theta_0 & = & \frac{1}{2}{\rm acos}\left[\frac{1+2{\sf h}^2 \cos
      2\theta_h}{\sqrt{1+4{\sf h}^4 + 4 {\sf h}^2\cos 2\theta_h}}\right].
\end{eqnarray}
with ${\sf h} = h/\sqrt{c_2\Delta E}$.  The angle $\theta_0$ obviously
gives the orientation of ${\bf \hat m}$.  Interestingly, it is an
analytic function of $h$ {\sl except} at $\theta_h=\pi/2$, i.e. when the
magnetic field is perpendicular to the $(100)$ axis.  As this value of
$\theta_h$ is approached, $\theta_0({\sf h})$ becomes sharper and
approaches a step function: $\theta_0({\sf h};\theta_h=\pi/2) =
\frac{\pi}{2}\Theta({\sf h}- 1/\sqrt{2})$.  It is also instructive to
plot the magnitude of the magnetization, $m({\sf h})$.  The
magnetization jumps at ${\sf h}=1/\sqrt{2}$ for $\theta_h=\pi/2$, but is
otherwise continuous (see Fig.~\ref{fig:magcurves}).

\begin{figure}[hbtp]
  \centering
  \includegraphics[width=0.45\textwidth]{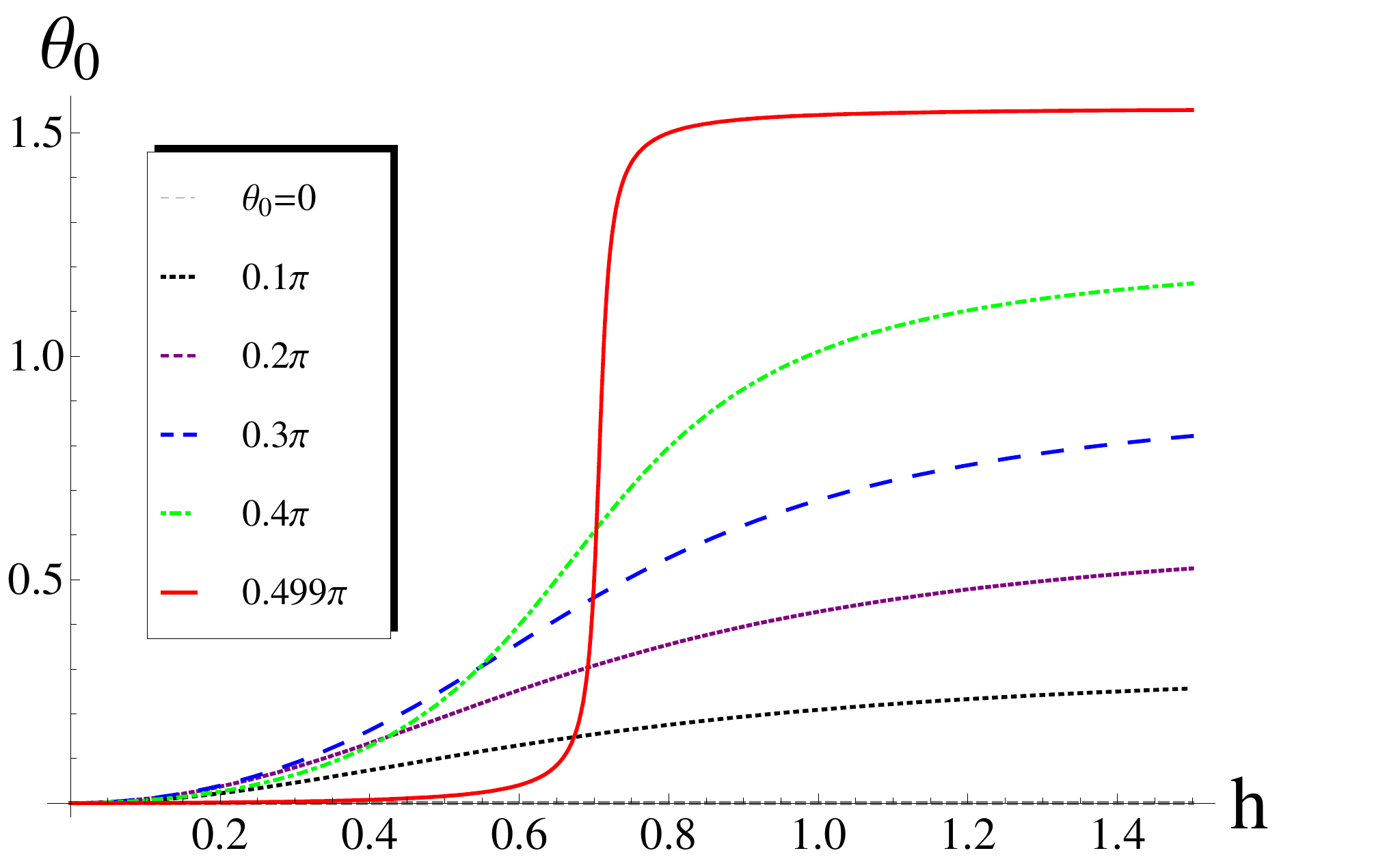}
  \caption{Ground state angle $\theta_0$ of the magnetization versus
    field ${\sf h}$ for $\theta_h=0,0.1\pi,0.2\pi,0.3\pi,0.4\pi,
    0.499\pi$ (from the bottom curve to the top curve).}
  \label{fig:theta0}
\end{figure}

Before ending this section, we comment on the range of validity of the
results.  First, though we have assumed throughout the above that
$c_1=0$, in fact it is possible to show that a spin flop (discontinuous
jump in the magnetization) occurs throughout region I of the phase
diagram in Fig.~\ref{c1-c2_phase_diagram}, in which $|c_1|<c_2$.  For
brevity, we do not give the (algebraically involved) argument here.
Second, we have assumed a particular ordering wavevector along the (110)
direction.  At zero field, this wavevector is chosen spontaneously from
among the family of equivalent $\langle$110$\rangle$ planes (e.g. (101)
etc.).  In the presence of a field, the different wavevectors will
become inequivalent, due to the magnetic anisotropy terms in $f_{110}$.
Hence, given enough time, annealing, or field cycling, the system may
choose the lowest free energy wavevector amongst this set in the
presence of the field.  This is rather clearly the wavevector which is
closest to the field axis.  In this situation, the situation
$\theta_h=\pi/2$ is avoided and the spin flop is avoided.  In practice,
wavevector reorientation is probably sufficiently slow at low
temperature to allow observation of the spin flop transition.

\section{Commensurability effects}
\label{sec:comm-effects}

Up to this point, our phenomenological theory leaves the {\sl phase} of
the spiral (i.e. the phase of ${\bf d}$) free.  In general, the
different directions within the spiral plane are not equivalent, and
when a fully account is taken of spin anisotropy and crystal symmetry,
the phase of the spiral may take preferred values.  In this section, we
discuss the effects of ``pinning'' of the phase, and how this leads to a
``lock-in'' transition for the spiral wavevector in some situations.

We will assume the spiral form in Eq.~(\ref{spiral_spin}), with some
given ${\bf q}(J_2/J_1)$ chosen to minimize the energy of the Heisenberg
Hamiltonian.  Using the arguments in the previous sections, we can fix
the {\sl plane} of the spiral, defined by the normal vector ${\bf\hat
  e}_3$.  Choosing two {\sl arbitrary} unit vectors spanning the plane
(${\bf \hat e}_1\times{\bf\hat e}_2 = {\bf\hat e}_3$), we can then write
\begin{equation}
 {\bf d}=m_s (\hat{\bf e}_1+i\hat{\bf e}_2)e^{i \theta}.
\label{eq:31}
\end{equation}
The terms considered up to now do not fix the phase $\theta$.  

The freedom to choose $\theta$ is related to translational invariance.
In particular, under a translation ${\bf r} \rightarrow {\bf r}+ {\bf
  a}$, we have 
\begin{equation}
  \label{eq:28}
  {\mathcal T}_{\bf a}: \qquad \theta \rightarrow \theta - {\bf q}\cdot
  {\bf a}.  
\end{equation}
Here ${\bf a}$ can be any Bravais lattice vector.  It is sufficient to
consider the primitive lattice vectors ${\bf
  a}=(0,\frac{1}{2},\frac{1}{2})$ and permutations.  We would like to
construct terms in the effective continuum Hamiltonian or Landau free
energy that are invariant under Eq.~(\ref{eq:28}), but which depend upon
$\theta$ directly and not only through its gradients.  Moreover, they must
also be periodic in $\theta$ (since a shift by $2\pi$ leaves ${\bf d}$
unchanged).  A general periodic functional of $\theta$ can be written
\begin{equation}
  \label{eq:30}
  V_\theta = -\sum_{n=1}^\infty \int\! d^3{\bf r}\, \lambda_n \cos
  (n\theta + \phi_n({\bf r})),
\end{equation}
where the $\lambda_n$ are arbitrary coefficients.  The $\phi_n$ are
arbitrary {\sl slowly varying functions} of ${\bf r}$, which should be
chosen, if possible, to ensure invariance under Eq.~(\ref{eq:28}).  The
functions should be slowly varying because large gradients of $\theta$
are heavily penalized by the Heisenberg Hamiltonian, which favors
constant $\theta$.  If $\theta$ varies slowly but $\phi_n$ varies
rapidly, then this term will average rapidly to zero on integration, and
can be neglected.

A general choice of function which achieves the desired invariance is
$\phi_n({\bf r}) = n{\bf q}\cdot {\bf r} ({\rm mod}\, 2\pi) +
\phi_{n0}$, with $\phi_{n0}$ a constant.  We need to determined which
(if any) of these functions is slowly varying.  This occurs if the
change in $\phi_n({\bf r})$ on shifting by a primitive lattice vector is
small.  By continuity, this is achieved when ${\bf q}$ is close to a
wavevector for which $\phi_n({\bf r})$ is constant under such a shift.
To achieve constancy, the $n^{\rm th}$ term should have $n{\bf
  q}\cdot {\bf a}$ a multiple of $2\pi$ for all three primitive vectors
${\bf a}$.  For this condition to hold for {\sl any} $n$, we require
that ${\bf q}\cdot {\bf a}$ be a rational multiple of $2\pi$.  We call
these special wavevectors satisfying this condition {\sl commensurate}.

Let us now specialize to a specific direction of wavevector of interest.
We take ${\bf q}=(q,q,0)$, corresponding to $J_2/J_1 \gtrsim 0.7$, which is
the case appropriate for MnSc$_2$S$_4$.  In this case, for the three
primitive translations, we have ${\bf q}\cdot {\bf a}=q/2,q/2,q$.  Thus
the condition for the wavevector to be commensurate is $n q/2 = 2\pi m$,
where $n,m$ are integers.   We assume the system is close to such a
value, i.e.
\begin{equation}
  \label{eq:32}
  q = 4 \,{\rm arccos}\, \left[ \sqrt{\frac{|J_1|}{8J_2}}\right]\approx
  q_{m,n}\equiv \frac{4\pi m}{n}.
\end{equation}
In general, the most important $m,n$ will be those with the smallest
$n$, since the terms $\lambda_n$ may be expected to decay with
increasing $n$.  For $J_2/J_1\gtrsim 0.7$ such that the $(q,q,0)$ order
is obtained, we find a number of commensurate wavevectors, shown in
Fig.~\ref{preferred_wave_vector}.  The smallest $q$ in this set is
$q=q_{3,8}=3\pi/2$, which is the wavevector observed in \MnScS.  The
presence of these other commensurate wavevectors with smaller $n$
suggests that other commensurate states might well be found by varying
$J_2/J_1$ by physical or chemical pressure.

Let us fix on the vincinity of one of these wavevectors.  Because the
other terms in $V_\theta$ rapidly oscillate, we need only keep the one
involving $q_{m,n}$:
\begin{equation}
  \label{eq:33}
   V_\theta = -\lambda \int\! d^3{\bf r}\, \cos
  (n\theta + n{\boldsymbol{ \delta q}}\cdot {\bf r} + \phi_{n0}),
\end{equation}
where ${\boldsymbol {\delta q}}= {\bf q}-{\bf q}_{m,n}$, with ${\bf
  q}_{m,n}=(q_{m,n},q_{m,n},0)$, and we simplified $\lambda_n
\rightarrow \lambda$.  This term favors configurations in which
$\boldsymbol{\nabla}\theta = - \boldsymbol{\delta q} - \phi_{n0}$, which
minimize the cosine.  Establishment of a phase gradient, however, costs
exchange energy.  This can be seen because from
Eqs.~(\ref{eq:31},\ref{spiral_spin}), a non-vanishing gradient
${\boldsymbol{\nabla}}\theta$ corresponds to a shift of wavevector.
Indeed, the physical wavevector ${\bf k}$ for general $\theta$ is
\begin{equation}
  \label{eq:34}
  {\bf k} = {\bf q}+{\boldsymbol\nabla}\theta.
\end{equation}
The exchange energy cost to distort the wavevector from ${\bf q}$ to
${\bf k}$ is, from Eq.~(\ref{eq:14})
\begin{equation}
  \label{eq:35}
  H_{ex} = \int \, d^3{\bf r}\,
  \frac{\kappa_{\mu\nu}}{2} \partial_\mu\theta \partial_\nu \theta,
\end{equation}
where the tensor stiffness $\kappa_{\mu\nu}$ is 
\begin{equation}
  \label{eq:38}
  {\bf \kappa} = \left( \begin{array}{ccc} \kappa_+ & \kappa_- & 0 \\
      \kappa_- & \kappa_+ & 0 \\ 0 & 0 & \kappa_3 \end{array}\right).
\end{equation}
Here $\kappa_1= \kappa_++\kappa_-$, $\kappa_2=\kappa_+-\kappa_-$, and
$\kappa_3$  are the stiffnesses along the principal axes.  At zero
temperature, they are given by:
\begin{eqnarray}
  \label{eq:36}
  \kappa_1 & = & \frac{J_1}{2}-\frac{J_1^2}{16J_2}+{\mathcal O}[J_3], \\
  \kappa_2 & = &  \frac{(8J_2-J_1)J_1 J_3}{16 J_2^2},\nonumber  \\
  \kappa_3 & = &  J_3 \frac{(128J_2^3-112 J_1 J_2^2 + 20 J_1^2 J_2 -
    J_1^3) }{8 J_1 J_2^2 }. \nonumber
\end{eqnarray}
For the most interesting case $q\approx 3\pi/2$, we have $\kappa_1
\approx (2+\sqrt{2})J_1/8$, $\kappa_2 \approx J_3/2$, and
$\kappa_3 \approx (\sqrt{2}+1)J_3$.  

We now proceed to analyze the effective Hamiltonian $H_{eff}=
H_{ex}+V_\theta$. Though we have given these expressions explicitly at
$T=0$, the general form in Eqs.~(\ref{eq:33},\ref{eq:35}) holds at any
temperature below the Neel temperature, with $H_{eff}$ replaced by
$F_{eff}$, the effective free energy, and with renormalized parameters
$\lambda(T), \kappa_i(T)$.  Moreover, because in this temperature range
the system exhibits magnetic long range order, the fluctuations of
$\theta$ are small and bounded, so that it is sufficient to consider
saddle points of the free energy.

It is convenient to shift variables to
$\tilde{\theta} = \theta+{\boldsymbol {\delta q}}\cdot {\bf r} +
\phi_{n0}/n$.  The free energy is
\begin{eqnarray}
  \label{eq:37}
  F_{eff} & = & \int\! d^3{\bf r}\, \big\{
  \frac{\kappa_{\mu\nu}}{2} \partial_\mu\tilde\theta \partial_\nu
  \tilde\theta + {\boldsymbol \delta}\cdot {\boldsymbol\nabla}\tilde\theta \nonumber
  \\
  && - \lambda \cos n\tilde\theta + \frac{1}{2} \delta q_\mu
  \kappa_{\mu\nu}\delta q_\nu\big\},
\end{eqnarray}
with $\delta_\mu = \kappa_{\mu\nu} \delta q_\nu$.  The last term is
independent of $\tilde\theta$ and can be neglected.  One can readily see
that
\begin{equation}
  \label{eq:42}
  {\bf k} = {\bf q}_{m,n} + {\boldsymbol\nabla}\tilde\theta.
\end{equation}
The minimum free
energy saddle points of $F_{eff}$ are translationally invariant along
the directions perpendicular to $\boldsymbol\delta$, which is along the
$(110)$ axis.  We therefore define the coordinate ${\sf x}=
(x+y)/\sqrt{2}$ along the $(110)$ direction, and rewrite the free energy
accordingly,
\begin{eqnarray}
  \label{eq:39}
  F_{eff} & = & A \int\! d{\sf x}\, \big\{ \frac{\kappa_1}{2} (\partial_{\sf
    x}\tilde\theta)^2 + \delta \partial_{\sf x}\tilde\theta - \lambda
  \cos n\tilde\theta \big\},
\end{eqnarray}
where $A$ is the area of the sample transverse to the $(110)$ axis,
$\delta = |{\boldsymbol\delta}| = q-q_{m,n}$, and we have dropped the
constant term.  It is now evident that $\delta$ enters only as a
boundary term, which means that the free energy depends upon $\delta$
only through the ``winding'' number $N_w = [\tilde\theta({\sf x}=L) -
\tilde\theta({\sf x}=0)]\frac{n}{2\pi}$ of the minimum energy saddle
point (across the length $L$ along the $(110)$ direction).  This allows
one to proceed by finding the saddle point energy for fixed $N_w$, and
then minimizing over $N_w$.

\begin{figure}[hbtp]
  \centering
    \includegraphics[width=0.45\textwidth]{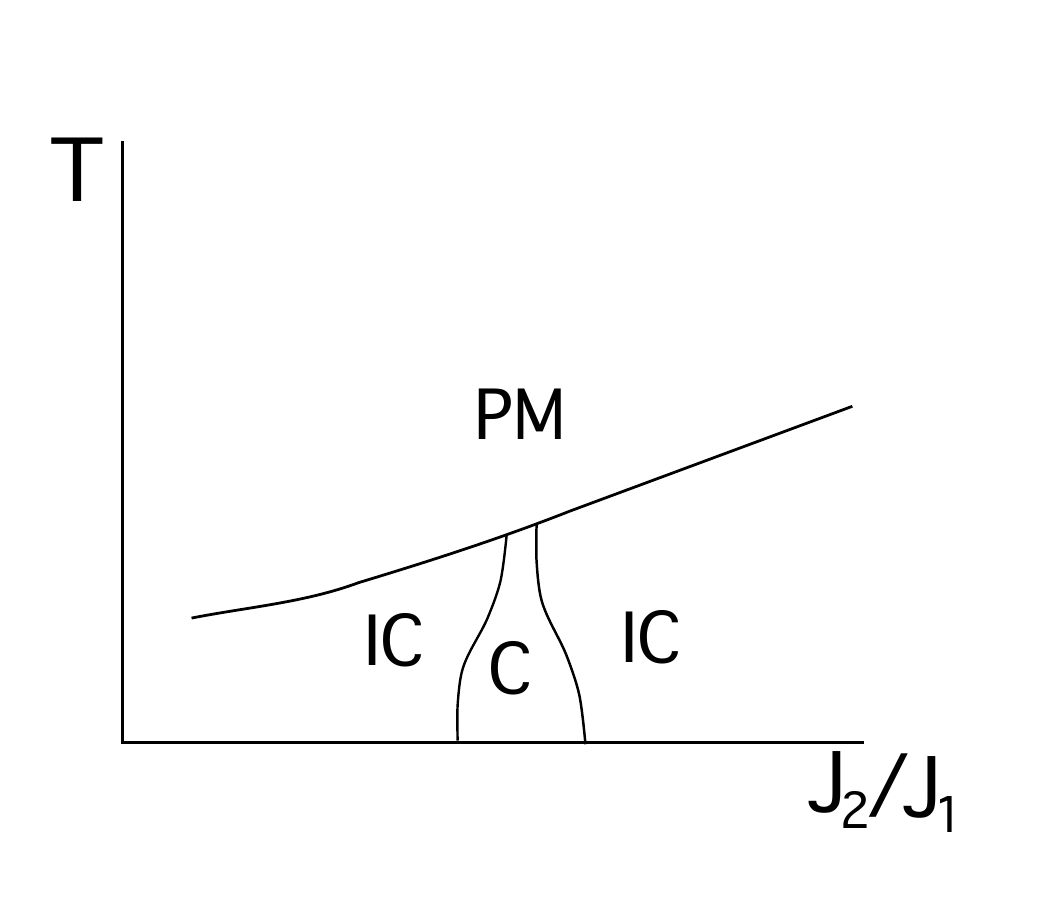}
  \caption{Schematic phase diagram showing commensurate (C) and
    incommensurate (IC) magnetic phases, and the paramagnetic  (PM)
    phase.  The figure is drawn as appropriate for a first order
    magnetic transition line, in which case the width of the
    commensurate phase remains non-zero on approaching the N\'eel
    temperature.} 
  \label{fig:CIT}
\end{figure}

It is useful to consider the cases $N_w=0$ and $N_w=\pm 1$.  For $N_w=0$,
the saddle point is uniform $\tilde\theta=0$ (up to a multiple of
$2\pi/n$).  For $N_w=\pm 1$, one has a single soliton solution:
\begin{equation}
  \label{eq:40}
  \tilde\theta({\sf x})=\frac{4}{n}\arctan\left[e^{\pm n
      \sqrt{\frac{\lambda}{\kappa_1}}({\sf x}-{\sf x}_0)}\right],
\end{equation}
where ${\sf x}_0$ is arbitrary and specifies the location of the center
of the soliton.  Note that the soliton width $w=
\frac{1}{n}\sqrt{\frac{\kappa_1}{\lambda}}$.  The energy of this
solution, for $\delta=0$, is $E_{N_w=1}-E_{N_w=0} = 8
\sqrt{\kappa_1\lambda}/n$.  When the spacing between solitons is much
larger than $w$, i.e. $L/|N_w| \gg w$, the energy of an $N_w$ soliton
state is approximately just $|N_w|$ times this single soliton energy.
Corrections to this non-interacting soliton approximation arise due to
the overlaps of the exponential tails of the solitons.  Defining the
mean soliton density as $n_w= N_w/L$, we may then write the
free energy density as
\begin{equation}
  \label{eq:41}
  f \sim \frac{8\sqrt{\kappa_1\lambda}}{n} |n_w| + \frac{2\pi\delta
    }{n} n_w + c |n_w| e^{-\frac{1}{w|n_w|}}, 
\end{equation}
where $c$ is a positive constant.  From Eq.~(\ref{eq:41}), the minimum
$n_w$ can be easily found.  For $|\delta|< |\delta_c|=4 \sqrt{\kappa_1\lambda}/\pi$,
one has $n_w=0$, and the wavevector is commensurate.  For
$|\delta|>|\delta_c|$, $n_w \neq 0$, and the wavevector becomes
incommensurate.   Due to fluctuations, one expects both $\lambda$ and
$\kappa_1$ to decrease with temperature.  Hence the width of the
commensurate state ($\propto |\delta_c|$) will decrease with increasing
temperature.  A schematic phase diagram is shown in Fig.~\ref{fig:CIT}.

We see that, when $J_2/J_1$ is close (but not too close!) to a value for
which the Heisenberg model alone has a commensurate spiral solution,
there is a ``lock-in'' transition on decreasing temperature from an
incommensurate to a commensurate spiral.  Within the commensurate (``C''
in Fig.~\ref{fig:CIT}) phase, the wavevector is constant and equal to
$q_{m,n}$.  This is consistent with observations on \MnScS.
Commensurate-Incommensurate transitions of this type are well studied,
and the reader interested in details of the associated critical behavior
may find it in various standard texts, for instance
Ref.~\onlinecite{ChaikinLubensky}.

%

\section{Quantum fluctuations}
\label{sec:quantum-fluctuations}

In this section, we develop a spin wave theory for the diamond
antiferromagnet, and obtain the leading quantum corrections to the spin
correlations.  

\subsection{Holstein-Primakoff bosons}
\label{sec:holst-prim-bosons}

We proceed in the standard way by defining Holstein-Primakoff bosons in
a spin coordinate frame rotated to follow the classical ordered state.
The local orthonormal axes will be defined by
\begin{eqnarray}
  \label{eq:43}
  {\bf\hat z}_i&=&{\hat {\bf S}}_i^{cl}={\rm Re}\, [{\bf d}e^{i{\bf q} \cdot {\bf
      r}_i}], \\
 {\bf \hat x}_i&=&-{\rm Im}\, [{\bf d}e^{i{\bf q} \cdot {\bf r}_i}], \nonumber\\
 {\bf\hat y}&=&-\frac{i}{2} {\bf d} \times {\bf d}^* = {\bf\hat e}_3 .\nonumber
\end{eqnarray}
Note that the ${\bf\hat y}$ axis is site independent, as it just
corresponds to the normal vector to the spiral plane.  The linearized
Holstein-Primakoff transformation is
\begin{equation}
  \label{eq:44}
   {\bf S}_i=(S-  n_i){\bf\hat z}_i+ \sqrt{2S}(a_i^{\dagger}\frac{({\bf\hat x}_i+i{\bf\hat y})}{2}+ a_i\frac{({\bf\hat x}_i-i{\bf\hat y})}{2}),
\end{equation}
which neglects corrections cubic in the canonical $a_i,a_i^\dagger$
boson operators.  Here $n_i = a^\dagger_i a^{\vphantom\dagger}_i$ as
usual.  It is convenient to pass from canonical bosons to ``coordinate''
and ``momentum'' operators,
\begin{equation}
  \label{eq:45}
  \chi_i=\dfrac{1}{\sqrt{2}}(a_i+a_i^{\dagger}), \hspace{1cm} 
 \xi_i=i\dfrac{1}{\sqrt{2}}(a_i^{\dagger}-a_i).
\end{equation}
The spin operator becomes
\begin{equation}
  \label{eq:46}
   {\bf S}_i=(S-  n_i){\bf\hat z}_i+\sqrt{S} \left(\chi_i {\bf\hat
       x}_i + \xi_i {\bf\hat y}\right),
\end{equation}
and
\begin{equation}
  \label{eq:48}
  n_i = \frac{\chi_i^2}{2}+\frac{\xi_i^2}{2} - \frac{1}{2}.
\end{equation}

\subsection{Spin wave Hamiltonian}
\label{sec:spin-wave-hamilt}

Inserting Eq.~(\ref{eq:46}) into the Heisenberg Hamiltonian, we obtain
terms of ${\mathcal O}(S^2), {\mathcal O}(S^{3/2})$, and ${\mathcal O}(S)$, dropping higher order
corrections:
\begin{eqnarray}
  \label{eq:47}
 H_{{\mathcal O}(S^2)} & = & \frac{1}{2} S^2 J_{ij} {\bf \hat{z}}_i \cdot {\bf
   \hat{z}}_j ,\\ 
H_{{\mathcal O}(S^{3/2})} & =& \frac{S \sqrt{S}}{2\sqrt{2}} J_{ij} {\bf \hat{z}}_i \cdot
{\bf \hat{x}}_j(\chi_i+\chi_j) , \nonumber \\
H_{{\mathcal O}(S)} & = & \frac{1}{2} S J_{ij} \big[(n_i+n_j) {\bf\hat{z}}_i
\cdot {\bf \hat{z}}_j + \chi_i\chi_j {\bf\hat{x}}_i
\cdot {\bf \hat{x}}_j + \xi_i \xi_j \big] .\nonumber
\end{eqnarray}

The ${\mathcal O}(S^{3/2})$ term vanishes because the local coordinate vectors are
eigenstates of the exchange matrix, e.g.
\begin{equation}
  \label{eq:49}
  J_{ij} {\bf\hat z}_j = J_m {\bf\hat z}_i,
\end{equation}
and ${\bf \hat x}_i \cdot {\bf \hat z}_i=0$.  Here $J_m$ is the minimum
eigenvalue of the exchange matrix.  The vanishing of the $O(S^{3/2})$
term is of course true because we expand about the classical ground
state.

Using Eq.~(\ref{eq:49}), one can further simplify the spin wave
Hamiltonian.  We obtain
\begin{eqnarray}
  \label{eq:50}
  && H_{{\mathcal O}(S)} = \\
&& - \sum_i \frac{SJ_m}{2} (\chi_i^2+\xi_i^2) +
  \sum_{ij}\frac{SJ_{ij}}{2} \left( \chi_i\chi_j {\bf\hat{x}}_i
\cdot {\bf \hat{x}}_j + \xi_i \xi_j \right),\nonumber
\end{eqnarray}
neglecting constant terms which do not affect the correlations.

\subsection{Action}
\label{sec:action}

Spin fluctuations are conveniently calculated using the path integral
approach.  The imaginary time action corresponding to Eq.~(\ref{eq:50})
has the usual Berry phase terms describing the canonical commutation
relations of $\chi_i,\xi_i$,
\begin{equation}
  \label{eq:51}
  S = \int_{\tau} \left\{H_{{\mathcal O}(S)} + \sum_i i \chi_{i} \partial_{\tau}
  \xi_{i} \right\}.
\end{equation}
Static correlations of $\chi_i$ and $\xi_j$ vanish, so we may consider
the two separately.  It is then convenient to integrate out one of these
fields to obtain an effective action for the other.  This gives
\begin{eqnarray}
  \label{eq:52}
  S_\chi & = & \frac{1}{2}\sum_{ij}\int_\tau \big\{ S \tilde{K}_{ij}
  \chi_i \chi_j + \frac{1}{S}[\tilde{J}^{-1}]_{ij} \partial_\tau
  \chi_i \partial_\tau \chi_j \big\}, \\
  S_\xi & = & \frac{1}{2}\sum_{ij}\int_\tau \big\{ S \tilde{J}_{ij}
  \xi_i \xi_j + \frac{1}{S}[\tilde{K}^{-1}]_{ij} \partial_\tau
  \xi_i \partial_\tau \xi_j \big\},
\end{eqnarray}
where
\begin{eqnarray}
  \label{eq:53}
  \tilde{J}_{ij} & = & J_{ij} - J_m \delta_{ij}, \\
  \tilde{K}_{ij} & = & J_{ij} {\bf\hat{x}}_i
\cdot {\bf \hat{x}}_j - J_m \delta_{ij}.
\end{eqnarray}

To diagonalize this, we move to momentum space.  Due to the sublattice
structure, we define two components for each field, $\chi_{A{\bf k}},
\chi_{B{\bf k}}$ and $\xi_{A{\bf k}},
\xi_{B{\bf k}}$, such that 
\begin{eqnarray}
  \label{eq:54}
  \chi_i & = & \int_{\bf k} \chi_{s(i){\bf k}} e^{i{\bf k}\cdot {\bf
      r}_i}, \\
  \xi_i & = & \int_{\bf k} \xi_{s(i){\bf k}} e^{i{\bf k}\cdot {\bf
      r}_i},
\end{eqnarray}
with $s(i) = A,B$ specifies the diamond sublattice of the site $i$. The
${\bf k}$ integral is defined as $\int_{\bf k} = v_{uc}\int\! \frac{d^3{\bf
  k}}{(2\pi)^3}$, where the integration domain is the first Brillouin zone, and
$v_{uc}=1/4$ is the volume of the real space unit cell.  It is
convenient to define 
\begin{equation}
  \label{eq:55}
  \hat\chi_{\bf k} = \left( \begin{array}{c} \chi_{A{\bf k}} \\
      \chi_{B{\bf k}} \end{array}\right), \qquad  \hat\xi_{\bf k} = \left( \begin{array}{c} \xi_{A{\bf k}} \\
      \xi_{B{\bf k}} \end{array}\right).
\end{equation}
The action becomes
\begin{eqnarray}
  \label{eq:56}
  S_\chi & = & \frac{1}{2}\int_{{\bf k}\omega} \hat\chi^T_{-{\bf
      k},-\omega} \cdot \overleftrightarrow{G}^{-1}_\chi({{\bf
      k},\omega}) \cdot \hat\chi_{{\bf
      k},\omega}, \\
  S_\xi & = & \frac{1}{2}\int_{{\bf k}\omega} \hat\xi^T_{-{\bf
      k},-\omega} \cdot  \overleftrightarrow{G}^{-1}_\xi({{\bf k},\omega})  \cdot \hat\xi_{{\bf k},\omega}.
\end{eqnarray}
Here the frequency integral is $\int_\omega = \int \!
\frac{d\omega}{2\pi}$ as usual.  The matrix Green's functions are
straightforwardly found, but somewhat cumbersome.  The reader interested
in the details is referred to Appendix~\ref{sec:spin-wave-greens}.  
With all these definitions, one can formally evaluate the equal time
correlation functions:
\begin{eqnarray}
  \label{eq:62}
  \langle \chi_i \chi_j \rangle & = & \int_{{\bf k},\omega}
  \left[G_\chi({\bf k},\omega)\right]_{s(j)s(i)} e^{i{\bf k}\cdot ({\bf
      r}_i - {\bf r}_j)}, \\
  \langle \xi_i \xi_j \rangle & = & \int_{{\bf k},\omega}
  \left[G_\xi({\bf k},\omega)\right]_{s(j)s(i)} e^{i{\bf k}\cdot ({\bf
      r}_i - {\bf r}_j)}. \nonumber
\end{eqnarray}
Here the subscripts give the matrix elements of the matrix Green's
functions. 

\subsection{Local moment}
\label{sec:local-moment}

\begin{figure}[t]
\vskip0.5cm
\scalebox{0.5}{\includegraphics{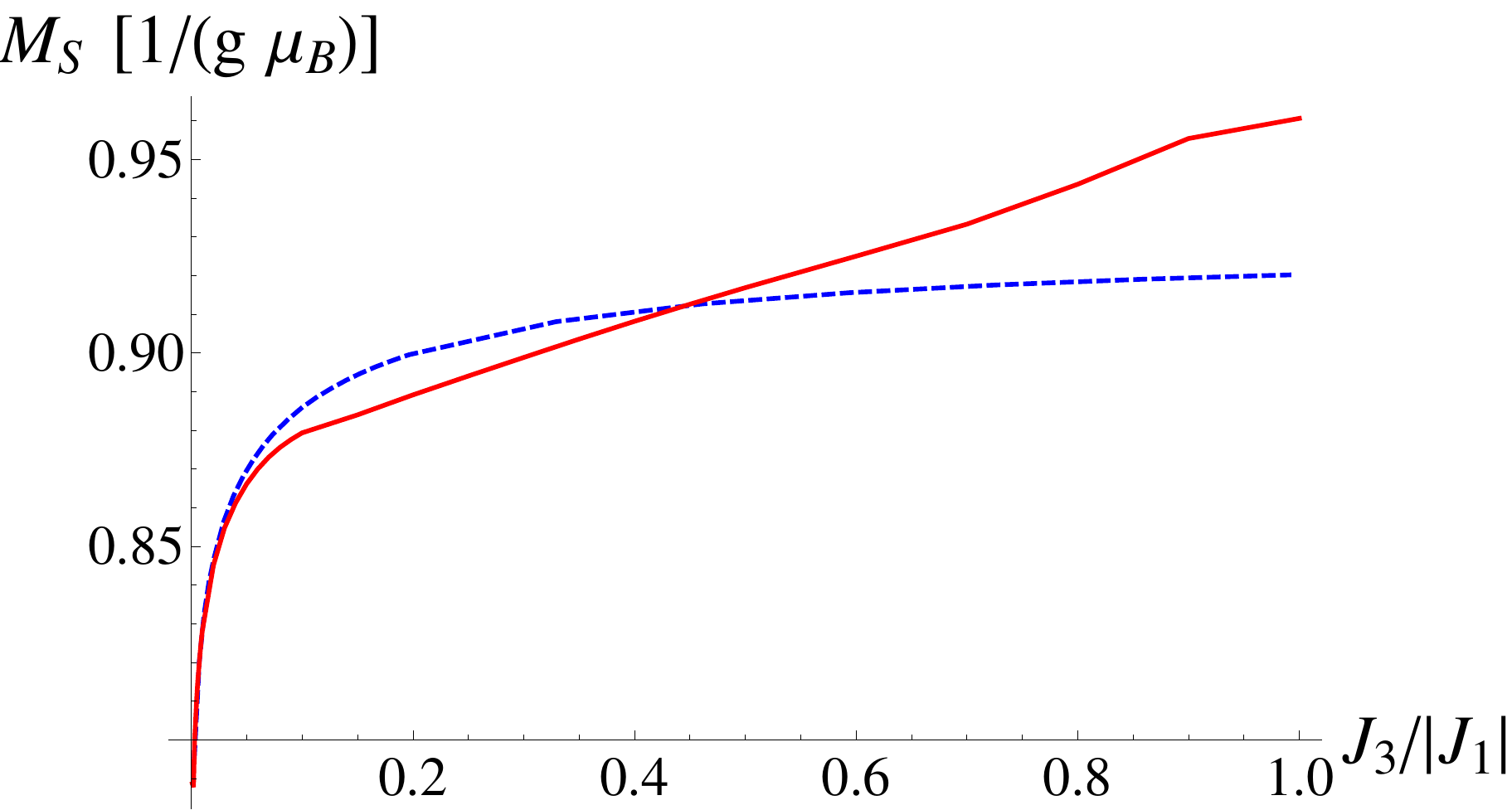}}
\caption{ Reduced magnetic moment $M_s$ as a function of
  $J_3/|J_1|$. The solid line is for fixed $J_2$, while the dashed line
  is for $J_2,J_3$ satisfying Eq.~\ref{eq:J2J3}, so that the wave vector
  remains equal to $3\pi/2(1,1,0)$.}
\label{reduced_mag}
\end{figure}

Focusing on the case of \MnScS, with ${\bf q}=(3\pi/2,3\pi/2,0)$, we
have calculated the reduction of the sublattice magnetization by
numerically evaluating the momentum integrals in Eqs.~(\ref{eq:62}) (the
frequency integration can be done analytically).  See Appendix~\ref{sec:frequency-integrals} for more details of the calculation.  The result for the on-site expectation value is
\begin{eqnarray}
  \label{eq:63}
  \langle \chi_i^2\rangle \approx 0.67, \qquad \langle \xi_i^2\rangle
  \approx 1.19,
\end{eqnarray}
for $J_3 = 0.1K \approx J_1/100$.  From this, one obtains $\langle
n_i\rangle \approx 0.43$ from Eq.~(\ref{eq:48}), which is approximately a
20$\%$ reduction from the classical local moment.   As $J_3$ is
increased, the moment increases closer to the classical value, as shown
in Fig.~\ref{reduced_mag}.
  
%
%

\section{Microscopic origin of magnetic anisotropy}
\label{sec:micr-orig-magn}

In
Sections~\ref{sec:magnetic-anisotropy},~\ref{sec:magn-proc},
we studied the effects of explicit spin rotation symmetry
breaking on {\sl phenomenological} grounds, using only the space group
symmetry of spinel structure. In this section, we address its
microscopic origins.  There are in general two mechanisms of spin
rotation symmetry violation in solids: (1) dipole interactions between
electron spins, and (2) spin-orbit coupling.  We consider both in turn,
and find these lead to somewhat different regimes of the
phenomenological model discussed previous.  Interestingly, only the
spin-orbit coupling mechanism can explain the observations in \MnScS.

\subsection{Dipolar interactions}
\label{sec:dipolar-interactions}

The dipole-dipole interaction can be written as
\begin{equation}
 {\bf H}_{D}=\dfrac{\mu_0}{4\pi} \sum_{i,j} \dfrac{{\bf m}_i \cdot {\bf m}_j}{r_{ij}^3} - \dfrac{3 {\bf m}_i \cdot {\bf r}_{ij} {\bf m}_j \cdot {\bf r}_{ij}}{r_{ij}^5},
 \label{dipole}
\end{equation}
where ${\bf m}_i = g \mu_B S {\bf S}_i$ is the dipole moment of the spin
$i$ (we included an explicit factor of $S$ to follow our convention of
unit vector spins).  Using $g\approx 2$ as expected for an $S=5/2$
Mn$^{2+}$ spin with a half-filled d shell, we obtain a dipolar energy of
interaction between two nearest-neighbor spins of approximately $0.5K$.
We note that this is not negligible (especially when added over many
spins within a correlation volume) but it is certainly weak compared to
the basic energy scale of exchange interactions as estimated from the
Curie-Weiss temperature $\Theta_{CW} \approx -23K$.  Therefore we expect
we can treat the dipolar interaction as a weak (but symmetry breaking)
perturbation on the ordered ground states of the Heisenberg model.  

\begin{figure}[hbtp]
  \centering
      \includegraphics[width=0.40\textwidth]{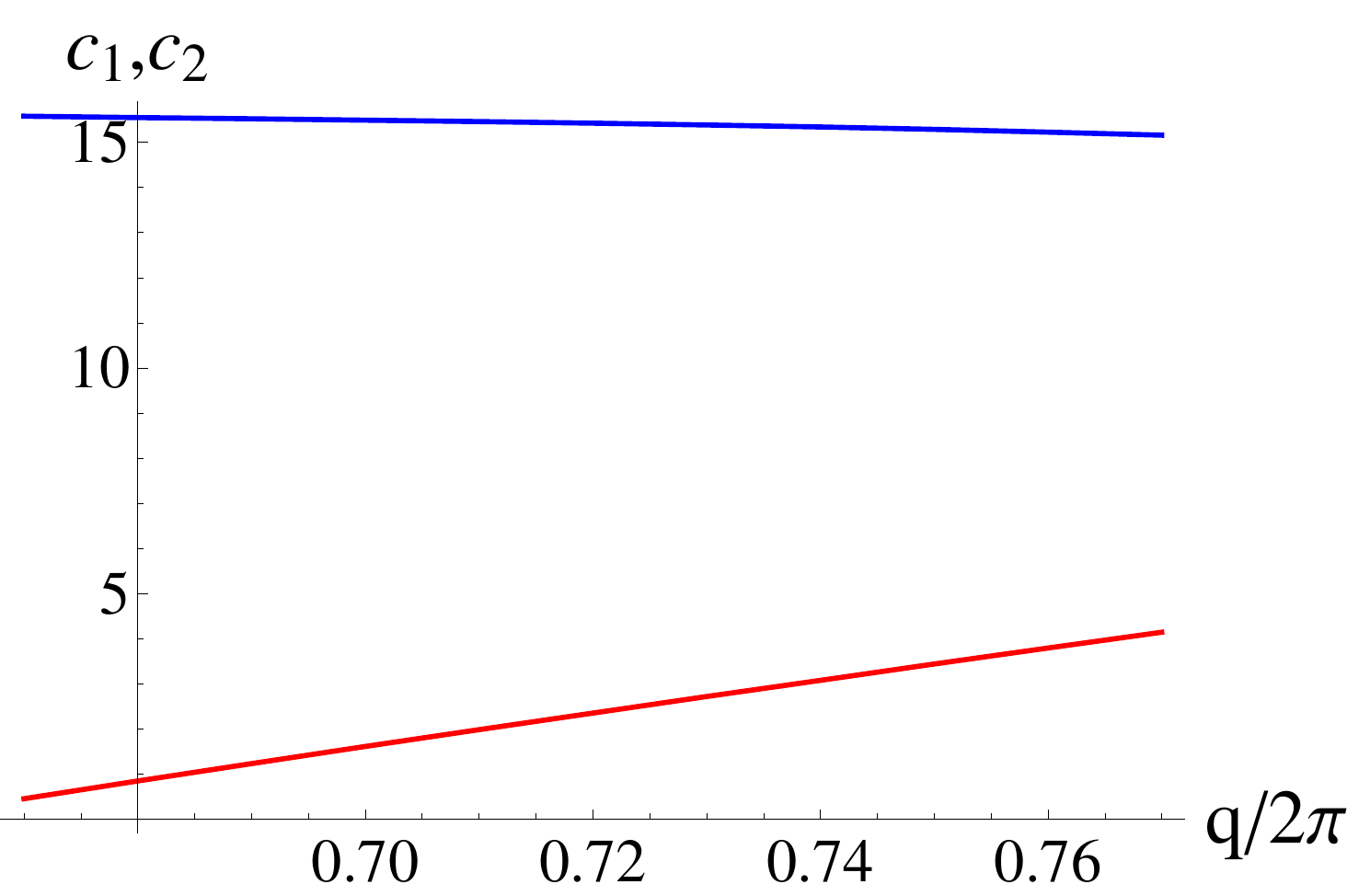}
      \caption{Calculated anisotropy parameters (in arbitrary units)
        $c_1$ (upper curve) and $c_2$ for ${\bf q}=(q,q,0)$ as a
        function of $q/(2\pi)$.}
  \label{fig:c1c2}
\end{figure}

To this end, we first consider the dipolar interaction classically by
simply inserting the general spiral form of Eq.~(\ref{eq:12}) in
Eq.~(\ref{dipole}) and evaluating the sum.  Because we are only
interested in the dependence of the energy upon the spin orientation of
the spiral, we may drop the first term in Eq.~(\ref{dipole}), which is
fully SU(2) invariant.  Because the spiral itself is at a non-zero
wavevector, there are no convergence difficulties with the long-range
dipolar sum.  Choosing the wavevector ${\bf q}=(q,q,0)$ as in
experiment, one indeed finds the form in Eq.~(\ref{non_vanishng_110}) is
obtained provided the sum is truncated in a manner preserving cubic
symmetry.  We plot the values of $c_1$ and $c_2$ in Eq.~(\ref{eq:24}) in
the physical range of $q$ for $0.7 \lesssim J_2/J_1<1$ in Fig.~\ref{fig:c1c2}.
Throughout this range we find $c_1>0$ and more than 3 times as large as $c_2$.
This favors alignment of spins within the plane normal to ${\bf\hat
  e}_3=(110)$.    Unfortunately, this is {\sl not} what is found
experimentally.  

Several possible complications should be considered before abandoning dipolar
interactions as a mechanism of magnetic anisotropy. First, in applying
Eq.~(\ref{dipole}) with ${\bf m}_i=g \mu_B S {\bf S}_i$, we have treated
the electron spins as point dipoles.  In fact, the electronic
wavefunctions may be somewhat extended.  Through such ``covalency'',
there may be some spin density not only in the atomic d orbital of the
Mn$^{2+}$ ion, but also on the neighboring chalcogenide p orbitals.
This can be approximately accounted for by modifying the dipole moment
distribution associated with a spin accordingly, to be distributed
amongst with a fractional moment $1-f$ on the central  Mn$^{2+}$ ion and
a fraction $f/4$ on each of the neighboring four $S^{2-}$ ions.  We
have carried out such a modified dipolar sum, and found that it does not
substantially alter the results of the point dipole model for a
reasonable range of parameters $f$.  

Another more interesting possibility is that fluctuations may alter the
dipolar energetics.  This is not an unreasonable possibility to consider
since, although the classical order parameter description is expected to
qualitatively (and indeed rather quantitatively) capture the long-range
order of the spins, the dipolar energy actually receives large
contributions from very nearby spins.  The latter could exhibit quite
different correlations from well separated spins which control the order
parameter.  

To consider this effect, we have calculated the leading corrections in
$1/S$ to the dipolar energy using the spin wave formalism described in
the previous chapter.  Since we treat the dipole-dipole interaction as a
perturbation, it is sufficient to consider the expectation value
$\langle H_D \rangle$ in each of the spin wave ground states specified
by ${\bf d}$.  To do so, we insert Eq.~(\ref{eq:46}) into $H_D$, and
expand to quadratic order in $\chi_i,\xi_i$, then take the expectation
value of the result.  The necessary correlators of $\chi_i,\xi_i$ are
calculated by numerical integration of Eq.~(\ref{eq:62}).  The values
obtained are given in Table~\ref{tab:swcorr}.  Because the basis vectors
${\bf\hat x}_i, {\bf\hat y}, {\bf \hat z}_i$ are expressed in terms of
${\bf d}$ in Eq.~(\ref{eq:43}), the result is again an energy function
of the form of Eq.~(\ref{eq:24}), which contains both the classical
expressions for $c_1,c_2$ and their leading quantum corrections.  We
find that the quantum corrections push the system even further from the
${\bf \hat e}_3=(110)$ state, and in any case the magnitude of the
corrections are very small compared to the classical values.  

\begin{table}[hbtp]
  \centering
  \begin{tabular}{c|c|c}
    ${\bf r}_{ij}$ & $\langle \chi_i \chi_j \rangle$ & $\langle \xi_i \xi_j
    \rangle$ \\ \hline
    {\bf 0} & 0.67 & 1.19 \\
    $\pm \tfrac{1}{2}(1,1,0)$ & 0.22 & -0.1 \\
    $\tfrac{1}{2}(\pm 1,0,\pm 1)$ & 0.18 & -0.3 \\
    $\tfrac{1}{4}(1,1,-1)$ & -0.25 & 0.23 \\
    $-\tfrac{1}{4}(1,1,1)$ & -0.25 & 0.23 \\
  \end{tabular}
  \caption{Numerically calculated values of correlations of
    $\xi_i$,$\chi_i$ fields from spin wave theory, for ${\bf
      q}=(3\pi/2,3\pi/2,0)$,  $J_2/J_1=\frac{1}{8}\cos^2(\pi/8)$, $J_3/J_1=0.01$.
    Values not specified have negligible correlations.} 
  \label{tab:swcorr}
\end{table}

Having thus exhausted the possible complications associated with the
dipolar interactions, we conclude that the observed ordered state in
\MnScS is inconsistent with a dipolar origin of the magnetic
anisotropy.  We therefore turn to spin-orbit effects in the following
subsection.

\subsection{Exchange anisotropy due to spin-orbit coupling}
\label{sec:anisotropic-exchange}

As we saw in the previous subsection, dipolar interactions do not appear
to be viable explanation of the orientation of the spin spiral observed
in \MnScS.  We now consider the second microscopic origin of magnetic
anisotropy, which is spin-orbit coupling.  From the point of view of
symmetry, the spinel lattice allows both single-ion (cubic) anisotropy
of the Mn$^{2+}$ spins and exchange anisotropy.  The former is however
expected to be extremely small for Mn$^{2+}$, which has an extremely
stable and isotropic 3d$^5$ configuration (one may expect a coupling
constant of a few {\sl millikelvin}).  However, exchange anisotropy is
non-negligible in many Mn magnets.  A microscopic calculation is beyond
the scope of this paper, but we can make a few statements on general
grounds.  Because of the closed shell configuration, these effects are
also expected to be much smaller than the typical exchange interactions
(i.e. perturbative in spin-orbit coupling).  However, they may still be
as large as or larger than the dipolar effects.  In \MnScS, one may
attempt to get some feeling for their magnitude by comparing the
measured effective moment seen in the Curie law $\mu_{eff} = 5.8\mu_B$
to the theoretical spin-only value $\mu_{S=5/2} = 2
\sqrt{\frac{5}{2}\frac{7}{2}} \approx 5.92$.  Given uncertainties in the
measurement, we expect no more than a $5-10\%$ deviation from the latter
(and very possibly a much better agreement masked by experimental
complications).  For Mn$^{2+}$, one expects that contributions to the
$g$-factor (which renormalize the effective moment) are {\sl second
  order} in the spin-orbit coupling.  Exchange anisotropy occurs at both
first order and second order.  At first order, one obtains the
antisymmetric Dzyaloshinskii-Moriya (DM) interaction, and at second
order symmetric exchange anisotropy.  Thus we would expect that the DM
interactions be of order $\sqrt{|\mu_{eff}-\mu_{5/2}|/\mu_{5/2}}
J_{ij}$ and symmetric exchange anisotropy be of order
$(|\mu_{eff}-\mu_{5/2}|/\mu_{5/2} )J_{ij}$.  

With this in mind, we consider the allowed {\sl form} of the exchange
anisotropy as constrained by the space group symmetry of the spinel
structure.  We first consider nearest-neighbor bonds.  Without loss of
generality, take a bond oriented along the $(111)$ axis.  DM interaction
is {\sl forbidden} on this bond because exactly between the two sites is
an inversion center (${\mathcal G}_5$ in Eq.~(\ref{symmetries})).  Thus
we need only consider exchange anisotropy.  This in turn is strongly
constrained by the $C_3$ rotation symmetry about the $(111)$ axis
(${\mathcal G}_3$ in Eq.~(\ref{symmetries})).  This allows only two
separate exchange couplings, for components parallel and perpendicular
to the bond, respectively.  We can write the associated exchange
Hamiltonian as 
\begin{eqnarray}
\label{eq:67}
H_{ani}^{n.n} &=&\sum_{\langle i,j \rangle} J_{\|} {\bf n}_{ij}\cdot
  {\bf S}_i \, {\bf n}_{ij}\cdot
  {\bf S}_j + J_{\bot} {\bf n}_{ij}\times
  {\bf S}_i \cdot {\bf n}_{ij}\times
  {\bf S}_j . \nonumber \\
\end{eqnarray}
There is a single parameter, $J_\perp-J_\parallel$, which parametrizes
the nearest neighbor exchange anisotropy.

Next, we consider the anisotropic exchange for next-nearest-neighbors.
Here the symmetry is considerably less constraining, since two second
neighbors (fcc neighbors) are not connected by a $C_3$ axis, and there
is no inversion center between them.  We have however determined the
most general exchange Hamiltonian between two such sites invariant under
all operations in Eq.~(\ref{symmetries}), which is a straightforward but
tedious calculation.  There is unfortunately no simple expression for
this Hamiltonian which describes all $6$ second neighbor bonds
simultaneously.  Instead we write the form for a particular bond,
connecting two sites $i,j$ on the ``A'' sublattice, separated by the
(arbitrarily chosen) Bravais lattice vector ${\bf
  r}_{ij}=(0,-\frac{1}{2},\frac{1}{2})$:
\begin{eqnarray}
\label{eq:68}
H_{ij}^{n.n.n}&=& J_a S_i^x S_j^x + D (S_i^x S_j^y -S_i^y S_j^x +S_i^x S_j^z -S_i^z S_j^x) \nonumber \\
&& + J_b (S_i^y S_j^z +S_i^z S_j^y)+J_c (S_i^y S_j^y + S_i^zS_j^z).
\end{eqnarray}
The full set of $H_{ij}$ for all other pairs of second neighbor sites
can be obtained by actions of symmetry operations on Eq.~(\ref{eq:68}),
which thus defines the full next nearest neighbor Hamiltonian
$H_{ani}^{n.n.n.}$.  Note that there are 3 symmetric exchange constants,
one linear combination of which represents the isotropic Heisenberg
term, and the other two ($J_b,J_a-J_c$) represent symmetric exchange
anisotropy.  Because of the absence of an inversion center between two
fcc sites in the spinel, there is an allowed DM term $D$.  However, the
presence of the inversion center implies that the $D$ term takes the
opposite sign for spins on the ``B'' sublattice.  

We can now consider the full exchange anisotropy Hamiltonian, $H_{ani}=
H_{ani}^{n.n} +H_{ani}^{n.n.n}-H_{\rm Heis.}$, as a perturbation to the
Heisenberg form, and evaluate the energy splittings induced for a given
spiral state specified by ${\bf q}$ and ${\bf d}$, by simply inserting
Eq.~(\ref{spiral_spin}) in $H_{\rm ani}$.  As required by symmetry, for
${\bf q}=(qq0)$ it again has the form of Eq.~(\ref{non_vanishng_110}).
Reading off the coupling constants, we find
\begin{eqnarray}
\label{eq:69}
    c_1
    &=&(J_a-J_c)(1+\sqrt{2})+(1-\frac{1}{\sqrt{2}})(J_{\perp}-J_{\parallel}), \nonumber \\
    c_2 &=& J_b . 
\end{eqnarray}
Note that the DM term $D$ does not enter these macroscopic anisotropy
parameters, which is a consequence of its staggered nature on the two
diamond sublattices.  

Unlike for the dipolar interactions, we see that Eqs.~(\ref{eq:69})
allow essentially arbitrary values of $c_1$ and $c_2$.  This means that,
in the absence of a microscopic calculation, the exchange anisotropy
mechanism is not inconsistent with the observed ordering in \MnScS,
which as we saw could be described phenomenologically by a range of
choices of $c_1,c_2$.  Given the {\sl incompatibility} of our dipolar
results, however, we tentatively conclude that spin-orbit induced
exchange anisotropy is likely at the origin of spin state selection in
\MnScS.

\section{Discussion}
\label{sec:discussion}

\subsection{Summary}
\label{sec:summary}

In this paper, we have extended the theory of
Ref.~\onlinecite{Doron:order_by_disorder} to describe the effects of
magnetic anisotropy and quantum fluctuations in frustrated
antiferromagnetic A-site spinels.  The theory predicts the possible
planes in which spins reside in the spiral magnetic ground states in
zero field, and describes their evolution with field.  In some
orientations a spin flop transition was found.  We described
commensurate-incommensurate transitions which occur below the N\'eel
temperature when the spiral wavevector locks to one of a set of specific
commensurate values.  These effects are all in accord with observations
on the best studied such material, \MnScS.  We addressed the reduced
static moment seen in \MnScS by spin wave calculations, and found that a
relatively large reduction can indeed be achieved by quantum
fluctuations due to the frustration-induced degeneracy, despite the
large $S=5/2$ spin of Mn$^{2+}$, if one assumes the third neighbor
exchange $J_3 \lesssim 0.1$.  Finally, we derived microscopic expressions
for the most important phenomenological magnetic anisotropy parameters,
taking into account both dipole-dipole interactions and spin-orbit
effects.  In \MnScS, we concluded the latter are most likely responsible
for the observed magnetic orientation.

\subsection{Experiments}
\label{sec:exper-impl}

Let us turn now to a further discussion of experiments.  First we
discuss existing results, and then consider future experiments. 

\subsubsection{local moment}
\label{sec:local-moment-1}

As mentioned above, from the weight in the magnetic Bragg peaks seen in
Ref.~\onlinecite{krimmel:magnetic_ordering} in \MnScS, it was estimated that
the local ordered moment $M_s \approx 0.8 M_{cl}$, where $M_{cl}$ is the
expected classical static moment for an $S=5/2$ spin.  In
Sec.~\ref{sec:local-moment}, we showed that the 17$\%$ reduction could
perhaps be due to quantum fluctuations, if $J_3$ is sufficiently small.
However, there are a number of reasons to be cautious about this
conclusion.  First, at a technical level, it is not clear to us how
large the experimental errors should be considered on this measurement,
which was done in a powder sample.  Second, the data was taken at
$T=1.5K$, more than half the ordering temperature $T_c=2.3K$, so thermal
fluctuations may contribute to some reduction of the moment.  

Finally, there are a number of different effects that have not been
addressed theoretically, which may contribute to the moment reduction.
First, we have neglected disorder, which is known to be present in the
form of inversion -- interchange of A and B site atoms of the spinel.
Such disorder can damage the spin spiral, reducing the {\sl ordered}
moment even if the local static moments remains large.  The nature of the
defects created and their impact on the ordered moment measured by
neutrons will be discussed in a separate future work.\cite{Savaryetal}\
A second effect that could contribute is a spin-orbit renormalization of
the $g$-factor.  Usually this is small in Mn$^{2+}$ magnets, but perhaps
this is something worth considering further.  

\subsubsection{microscopics of anisotropy}
\label{sec:micr-anis}

As discussed above and in Sec.~\ref{sec:micr-orig-magn}, though dipolar
interactions between Mn$^{2+}$ spins might seem a likely candidate for
the origin of the magnetic anisotropy in \MnScS, they appear to be
inconsistent with the observed nature of this anisotropy.  While we can
reconcile the existing experiments with a picture of spin-orbit induced
anisotropy (with some assumptions), it is still surprising to us that
such effects would be competitive with dipolar interactions.  We
believe the conflict of the latter with the ordered state seen in \MnScS
is a significant one, and found in Sec.~\ref{sec:dipolar-interactions}
that neither covalency nor quantum fluctuations were likely to effect a
reconciliation.  

One possibility we have {\sl not} considered is the effect of disorder
and granularity.  Given the long-range nature of the dipolar
interaction, it is possible that defects created by disorder in an ideal
spiral can facilitate large changes in the dipolar energy.  This is an
interesting issue to be explored in the future.  We emphasize that,
although such a mechanism of anisotropy might be possible, the
phenomenological portion of our theory is entirely independent of these
details and is quite generally valid irrespective of the microscopic
physics of anisotropy. 

\subsubsection{magnetization experiments}
\label{sec:magn-exper}

We now turn to future experiments.  Of particular interest would be the
development of single crystals.  This has already been emphasized in
Ref.~\onlinecite{Doron:order_by_disorder}, where predictions were made for unusual
``spiral surface'' structure in the angle-resolved neutron structure
factor.  Based on the results of this paper, we suggest that
single crystals are also interesting for the study of magnetization
effects.  An obvious suggestion is to look for signs of the spin flop
transition discussed in Sec.~\ref{sec:anisotropic-case}.  Another
interesting measurement would be torque magnetometry.  As shown in
Fig.~\ref{fig:theta0}, the angle of the magnetization can be strongly
misaligned with the applied field, which should leads to a large torque.
This is a very sensitive technique that perhaps does not require as
large crystals as neutron scattering does.

\subsection{Ferroelectricity}
\label{sec:ferroelectricity}

Our results enable us to discuss magnetically-induced ferroelectricity
in the A-site spinels.  This may be expected since many recent studies,
both theoretical and experimental, have emphasized the relation between
spiral spin states and ferroelectricity.  The basis for such a
relationship goes back much earlier to symmetry considerations of Landau
\cite{landau} and Dzyaloshinskii.\cite{dzyaloshinskii} Several recent
studies have pointed out that very general arguments suggest a simple
relationship 
between the electric polarization {\bf P} and the basic parameters
${\bf\hat e}_3$ and ${\bf q}$ describing the spiral:
\cite{kenzelmann:magnetic_inversion_symmetry_breaking_ferroelectricity}
\cite{ferroelectricity_in_spiral_magnets}
\cite{aharonov:topologica_quantum_effect}
\begin{equation}
\label{eq:70}
 {\bf P} \propto {\bf e}_3 \times {\bf q}.
\end{equation}
Here, as in the text, ${\bf\hat e}_3$ is the axis which is perpendicular
to the plane of the spins and {\bf q} is the wavevector.  

The argument leading to Eq.~(\ref{eq:70}) is rather simplified, and
actually assumes a sort of ``spherical symmetry''.  In reality, in the
reduced crystal symmetry environment of the solid, the actual relation may be
somewhat different.  Still, for the A-site spinels, a complete symmetry
analysis leads to rather similar results.  In particular, time reversal
symmetry allows a quadratic term in the ${\bf d}$ order parameter
(which is time reversal odd) to couple linearly to ${\bf P}$.  One
therefore expects the polarization to take the form
\begin{equation}
  \label{eq:71}
  P_\alpha = c_{\alpha\beta\gamma}({\bf q}) d_\beta^* d_\gamma^{\vphantom*}.
\end{equation}
As argued earlier, all such bilinears in ${\bf d}$ can be rewritten in
terms of ${\bf\hat e}_3$.  The coefficients $c_{\alpha\beta\gamma}$ are
constrained by crystal symmetry.  Specifically, we require that the left
and right hand sides of Eq.~(\ref{eq:71}) transform identically under
the little group which leaves ${\bf q}$ invariant.  

For ${\bf q}=(q,q,q)$, applying Eqs.~(\ref{eq:2}), we find the
form 
\begin{equation}
  \label{eq:72}
  {\bf P}_{111} = c_1 \begin{pmatrix} e_3^x \\ e_3^y \\ e_3^z \end{pmatrix}+
  c_2 \begin{pmatrix} e_3^z \\ e_3^x \\ e_3^y \end{pmatrix}+
  c_3 \begin{pmatrix} e_3^y \\ e_3^z \\ e_3^x \end{pmatrix} .
\end{equation}
The simplified Eq.~(\ref{eq:70}) corresponds to $c_1=0$, $c_3=-c_2$.
However, in general, symmetry allows any values of $c_1,c_2,c_3$.  

For ${\bf q}=(q,q,0)$, using Eqs.~(\ref{eq:8}), we find instead
\begin{equation}
  \label{eq:73}
{\bf P}_{110} = c_1 \begin{pmatrix} e_3^z \\ -e_3^z \\ 0 \end{pmatrix}+
  c_2 \begin{pmatrix} 0 \\ 0 \\ e_3^x-e_3^y \end{pmatrix}.
\end{equation}
Eq.~(\ref{eq:70}) is the special case $c_2=-c_1$.

Given these results, we can make some limited predictions on the
ferroelectric polarization in the A-site spinels.  In \MnScS, where the
ordering wavevector and spiral plane is known, we can directly apply
Eq.~(\ref{eq:73}) without much ambiguity.  We have ${\bf\hat e}_3= {\bf
  \hat z}$, which means that there is a spontaneous polarization with
${\bf P}$ along the $1\bar{1}0$ direction.  It would be
interesting to search for this experimentally in single crystals, or for
dielectric anomalies related to this in powders.  Moreover, the
phenomenological theory in Sec.~\ref{sec:anisotropic-case}, in
conjunction with Eq.~(\ref{eq:73}), describes how this polarization may
be rotated by an applied field.  Again, detailed single crystal studies
would be enlightening.

For spinels in the regime where ${\bf q}=(q,q,q)$, the theory is
somewhat less predictive.  This is because not only is there ambiguity
in the spiral plane giving ${\bf\hat e}_3$ (due to the unknown constant
$c$ in Eq.~(\ref{eq:11})), but also there are more unknowns in the
relation between the polarization and the spiral plane,
Eq.~(\ref{eq:72}).  A microscopic theory for Eq.~(\ref{eq:72}), which
determines the $c_i$, is therefore desirable.  We imagine one might be
constructed based on the inverse Dzyaloshinskii-Moriya
interaction mechanism\cite{ferroelectricity_in_spiral_magnets},
since we have seen that there is a single DM interaction allowed in the
A-site spinels -- see Eq.~(\ref{eq:68}).  The  polarization
can be very sensitive to details of the microscopics.  For instance, for
$c>0$ in Eq.~(\ref{eq:11}), we have ${\bf \hat e}_3 = (1,1,1)/\sqrt{3}$,
and according to  Eq.~(\ref{eq:70}), the polarization {\sl
  vanishes}.  However, in general this is an artifact of the
simplifications in Eq.~(\ref{eq:70}), and according to
Eq.~(\ref{eq:72}), ${\bf P}\neq {\bf 0}$.  However the orientation of
the polarization is precisely controlled by deviations from the na\"ive
Eq.~(\ref{eq:70}).

\begin{acknowledgments} 
 
  The authors would like to thank A. Loidl, A. Krimmel, and M. M\"ucksch
  for discussions and correspondence.  This research was supported by
  the Packard Foundation and the National Science Foundation through
  \mbox{grant DMR04-57440}.
  
\end{acknowledgments}  

\appendix

\section{Splitting of spiral surface degeneracy}
\label{sec:splitt-spir-surf}

\begin{figure}[hbtp]
  \centering
  \includegraphics[width=0.45\textwidth]{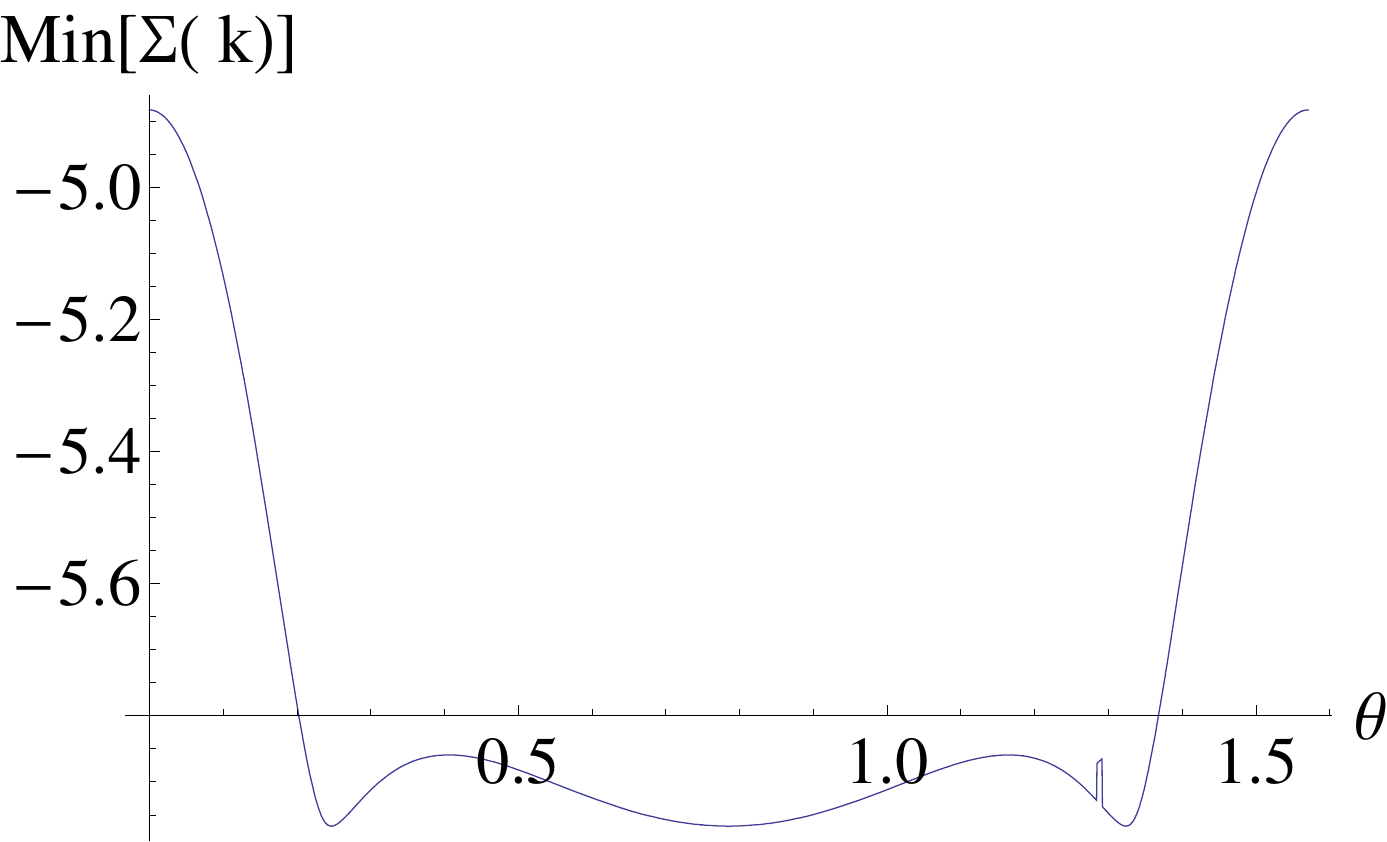}
  \caption{Minimum value of $\Sigma({\bf k})$ for ${\bf k}$ of the form
    ${\bf k}=(q\cos\theta,q\sin\theta,k)$ as a function of $\theta$, for
    $J_2/J_1=0.4$.  For {\sl all} values of $J_2/J_1$, the minimum value
    is achieved at $\theta=\pi/4$.} 
  \label{fig:mintheta}
\end{figure}

In this appendix, we give some details on how the ground state spirals
are determined in the presence of third neighbor antiferromagnetic
exchange $J_3$.  First, we performed a numerical study of the minima of
Eq.~(\ref{eq:6}), considering only wavevectors fixed on the spiral
surface, i.e. satisfying $\Lambda({\bf k}) = \lambda = 1/8J_2$.  These can be
conveniently studied by solving this condition to give $k_z$ in terms of
$k_x,k_y$:
\begin{equation}
  \label{eq:59}
  k_z = \pm 4 \arccos \left[ \left( \frac{\lambda^2 - \sin^2\frac{k_x}{4}
        \sin^2\frac{k_y}{4}}{\cos^2\frac{k_x}{4}\cos^2\frac{k_y}{4} - \sin^2\frac{k_x}{4}
        \sin^2\frac{k_y}{4}}\right)^{1/2}\right].
\end{equation}
Here the solution (and the surface) exists only when the argument of the
square root is between 0 and 1.  Inserting this value of $k_z$ into
Eq.~(\ref{eq:6}), we can obtain the energy on the surface explicitly.
One can then scan linearly along lines defined by ${\bf
  k}=(q\cos\theta,q\sin\theta, k)$ on the surface and determine the
lowest energy for each $\theta$.  In every case, the lowest energy as a
function of $\theta$ is achieved for $\theta=\pi/4$ (an example is shown
in Fig.~\ref{fig:mintheta}), which implies a
wavevector of the form $(q,q,k)$ on the surface.

Having determined that the ground state wavevector is always of the
$(q,q,k)$ form, we need only search this ray for the ground state.  This
can be done analytically.  One obtains
\begin{eqnarray}
  \label{eq:64}
  &&  \Sigma(q)\equiv \Sigma(q,q,k_z(q)) \nonumber \\
&& =  \frac{1}{64} \Big[96 (8 \lambda^2-3) \lambda^2+(256 \lambda^4-13) \cos
  q \nonumber \\
  && +2 (16 \lambda^2+5) \cos 2 q-3 \cos 3 q+6\Big] \sec^2\frac{q}{2} .
\end{eqnarray}
This should be evaluated only when such a wavevector exists on the
surface.  This condition is
\begin{eqnarray}
  \label{eq:65}
  && \sin^2 \frac{q}{4} < \lambda \qquad \mbox{or}\qquad
  \sin^2\frac{q}{4} > 1-\lambda.
\end{eqnarray}

Now it is simple to study the ground states.  One can check that the
$(q,q,q)$ state, for which $q=\arccos[(8\lambda^2-5)/3]$, is always a
local minimum of Eq.~(\ref{eq:64}).  It however only exists however when
this value is well-defined, which requires $\lambda > 1/2$.  This
corresponds to $1/4<J_2<1/2$.  Indeed, in this range it is
straightforward to show that this is the global energy minimum.

For $J_2$ sufficiently large, one can readily see that the minimum of
Eq.~(\ref{eq:64}) is instead achieved at the boundary of its domain of
validity, i.e. when the inequalities in Eq.~(\ref{eq:65}) are satisfied
as {\sl equalities}.  This corresponds to $k_z(q)=0$, i.e. a $(q,q,0)$
state.  This eventually ceases to be a minimum for small enough $J_2$.
A choice of such wavevector is $q=q_0 = 4\arcsin \sqrt{\lambda}$.  For
this to be a minimum, we need $\Sigma'(q_0)<0$.  By differentiating
Eq.~(\ref{eq:64}) and evaluating, we find
\begin{equation}
  \label{eq:66}
  \Sigma'(q_0) = 16 \left[\lambda(1-\lambda)\right]^{3/2}
    \frac{1-6\lambda+4\lambda^2}{2\lambda-1}. 
\end{equation}
It is straightforward to show that this is negative provided $\lambda<
(3-\sqrt{5})/4$ or $J_2> 1/[2(3-\sqrt{5})]$, which determined the domain
of the $(q,q,0)$ state.  In between this and the $(q,q,q)$ state we
necessarily have the $(q,q,q^*)$ state.

\section{Spin wave Green's functions}
\label{sec:spin-wave-greens}

In this appendix, we give some details of the spin wave Green's
functions.  The Green's functions defined in Eq.~(\ref{eq:56}) can be
written as
\begin{eqnarray}
  \label{eq:57}
  \overleftrightarrow{G}_\chi({\bf k},\omega) & = & \left[ S \overleftrightarrow{B}({\bf k}) +
    \omega^2 (S \overleftrightarrow{A}({\bf k}))^{-1}\right]^{-1}, \\
  \overleftrightarrow{G}_\xi({\bf k},\omega) & = & \left[ S \overleftrightarrow{A}({\bf k}) +
    \omega^2 (S \overleftrightarrow{B}({\bf k}))^{-1}\right]^{-1}.
\end{eqnarray}    
Here we've defined a number of matrices occurring as Fourier transforms
of exchange matrices:

\begin{eqnarray}
\label{eq:58}
\overleftrightarrow{A}({\bf k})&=&\overleftrightarrow{W}_{{\bf q},\gamma}({\bf k}), \qquad \overleftrightarrow{B}({\bf k})=\overleftrightarrow{W}_{{\bf
  0},0}({\bf k}), \\ 
\overleftrightarrow{W}_{\bf k^{\prime}}({\bf k})&\equiv&\left(
    \begin{array}{cc}
    W_{{\bf k}^{\prime},\gamma}^{11}({\bf k}) & W_{{\bf k}^{\prime},\gamma}^{12}({\bf k}) \\
    W_{{\bf k}^{\prime},\gamma}^{21}({\bf k}) & W_{{\bf k}^{\prime},\gamma}^{22}({\bf k})
    \end{array}
    \right).
\end{eqnarray}
The elements of $\overleftrightarrow{W}$ are conveniently given in terms
of the nearest-neighbor vectors ${\bf n}_a$ of the A sites of the
diamond lattice,
\begin{eqnarray}
  \label{eq:61}
  {\bf n}_0 & = & \tfrac{1}{4}(1,1,1), \qquad
  {\bf n}_1  =   \tfrac{1}{4}(1,-1,-1), \\
  {\bf n}_2 & = & \tfrac{1}{4}(-1,1,-1), \qquad
  {\bf n}_3  =  \tfrac{1}{4}(-1,-1,1).
\end{eqnarray}
Then
\begin{eqnarray}
\label{eq:60}
W_{{\bf k}^{\prime},\gamma}^{11}({\bf k}) &=& W^{22}_{{\bf k}^{\prime},\gamma}({\bf k}) 
\\
&= & -J_{m}+ J_2\sum_{a\neq b}  e^{i {\bf k} \cdot ({\bf n}_a-{\bf n}_b)}
\cos{{\bf k}^{\prime} \cdot ({\bf n}_a-{\bf n}_b)}, \nonumber \\  
W_{{\bf k}^{\prime},\gamma}^{12}({\bf k}) &=& \left[W_{{\bf k}^{\prime},\gamma}^{21}({\bf
    k})\right]^* \nonumber \\
& = &  J_1\sum_a e^{i {\bf k}\cdot {\bf n}_a} \cos({\bf k^{\prime}} \cdot {\bf n}_{a}+\gamma) \\
&& \hspace{-0.4in} + \frac{1}{2}J_3\!\!\sum_{a\neq b \neq c \neq a}\!\! e^{i {\bf k}\cdot ({\bf n}_a+{\bf
    n}_b-{\bf n}_c)} \cos[{\bf k^{\prime}} \!\cdot\!({\bf n}_a\!+\!{\bf
  n}_b\!-\!{\bf n}_c)\!+\!\gamma]
.\nonumber 
\end{eqnarray}
Here the sums range over distinct values of $a,b,c$ taken from
$0,1,2,3$.  

\section{Frequency integrals of Green's functions and momentum integration in brillouin zone}
\label{sec:frequency-integrals}

In this appendix, we give some details of the frequency integrals of Green's functions and the transformation to unit variables $x_i$ in momentum space.
The frequency integrals of the correlation functions defined in Eq.~(\ref{eq:62}) can be calculated analytically using the following relations.

\begin{eqnarray}
\label{eq:freq_int}
\int_{\omega} \frac{1}{{\omega}^4+p_1{\omega}^2+p_2} &=& \frac{1}{2\sqrt{p_1}\sqrt{p_1+2\sqrt{p_2}}} \nonumber \\
\int_{\omega} \frac{{\omega}^2}{{\omega}^4+p_1{\omega}^2+p_2} &=& \frac{1}{2\sqrt{p_1+2\sqrt{p_2}}} \nonumber 
\end{eqnarray}

The frequency integrated Green's functions, $\overleftrightarrow{G}_{\chi (\xi)}({\bf k})$, can be written as 

\begin{eqnarray}
\label{eq:int_green}
\overleftrightarrow{G}_{\chi (\xi)}({\bf k}) &\equiv& \left(
     \begin{array}{cc}
     G_{\chi (\xi)}^{11}({\bf k}) &  G_{\chi (\xi)}^{12}({\bf k})\\
     G_{\chi (\xi)}^{21}({\bf k}) & G_{\chi (\xi)}^{22}({\bf k})
     \end{array}
     \right)
\end{eqnarray}

with

\begin{eqnarray}
\label{eq:green_element}
G_{\chi}^{11}({\bf k})&=& G_{\chi}^{22}({\bf k}) \\
&=& \frac{1}{C({\bf k})}(\overleftrightarrow{B}^{11}({\bf k})+D({\bf k}) \overleftrightarrow{A}^{11}({\bf k})) \nonumber \\
G_{\chi}^{12}({\bf k})&=& [G_{\chi}^{21}({\bf k})]^*  \\
&=& \frac{1}{C({\bf k})}(\overleftrightarrow{B}^{12}({\bf k})-D({\bf k}) \overleftrightarrow{A}^{12}({\bf k})) \nonumber \\
\end{eqnarray}

Here $\overleftrightarrow{A}^{\alpha}({\bf k})$ is the $\alpha$ matrix element of $\overleftrightarrow{A}({\bf k})$ defined in Eq.~\ref{eq:58}. 
Then, C({\bf k}),D({\bf k}) are,

\begin{eqnarray}
C({\bf k})&\equiv& 2\sqrt{Tr[{\overleftrightarrow{A}({\bf k})\overleftrightarrow{B}({\bf k})}]+2|{\overleftrightarrow{A}({\bf k})\overleftrightarrow{B}({\bf k})}|} \nonumber \\
D({\bf k}) &\equiv& \sqrt{\frac{|\overleftrightarrow{B}({\bf k})|}{|\overleftrightarrow{A}({\bf k})|}} \nonumber 
\end{eqnarray}

It is natural from Eq.~\ref{eq:57} that $\overleftrightarrow{G}_{\xi}({\bf k})$ can be expressed $\overleftrightarrow{G}_{\chi}({\bf k})$ with the changes $\overleftrightarrow{A}({\bf k}) \leftrightarrow \overleftrightarrow{B}({\bf k})$

The numerical integration of the momentum in the 1st brillouin zone, can be eqsily evaluated using the transformation to the unit variables in momentum space.
\begin{eqnarray}
\label{eq:trans_variables}
{\bf k} &=& {\bf b}_1 x_1+ {\bf b}_2 x_2 +{\bf b}_3 x_3 \nonumber \\
{\bf b}_i &=& \frac{2\pi  {\bf a}_j \times {\bf a}_k}{{\bf a}_i \cdot ({\bf a}_j \times {\bf a}_k)} \nonumber 
\end{eqnarray}

where ${\bf a}_i$ is the fcc primitive vectors, permutations of 1/2(0,1,1).
Hence we can transform the momentum {\bf k} to unit variables $x_i$, then the momentum integration in the 1st brillouin zone can be written as
\begin{eqnarray}
\label{eq:unit_var_int}
v_{uc}\int_{BZ}\! \frac{d^3{\bf
  k}}{(2\pi)^3} 
  & \rightarrow &  \displaystyle\prod_{i=1}^3 [\int_{0}^{1} dx_i]
\end{eqnarray}


\vfill\eject

\end{document}